\documentclass{sig-alternate-2013}

\usepackage{amssymb,amsfonts,amsmath,amscd,dsfont,mathrsfs}
\usepackage{graphicx,float,psfrag,epsfig}
\usepackage{wrapfig}
\usepackage[utf8]{inputenc}
\usepackage{epstopdf,color}
\usepackage{hyperref,bbm}
\usepackage{multirow}
\usepackage{multicol}
\usepackage{algorithm}
\usepackage[noend]{algpseudocode}
\usepackage{balance}
\usepackage{caption}
\usepackage{flushend}
\usepackage{kbordermatrix}

\usepackage{breqn}
\makeatletter
\def\BState{\State\hskip-\ALG@thistlm}
\makeatother

\let\oldReturn\Return
\renewcommand{\Return}{\State\oldReturn}

\DeclareMathAlphabet{\mathpzc}{OT1}{pzc}{m}{it}

\newtheorem{propo}{Proposition}[section]
\newtheorem{lemma}[propo]{Lemma}

\newtheorem{corollary}[propo]{Corollary}
\newtheorem{theorem}[propo]{Theorem}

\newtheorem{claim}[propo]{Claim}

\newtheorem{innercustomthm}{Theorem}
\newenvironment{customthm}[1]
  {\renewcommand\theinnercustomthm{#1}\innercustomthm}
  {\endinnercustomthm}

\newcommand{\reals}{{\mathbb R}}
\newcommand{\prob}{\mathbb P}

\newcommand{\hv}{\hat{v}}
\newcommand{\vs}{v^*}

\newcommand{\adt}{{\alpha_{d,\theta}}}
\newcommand{\lT}{{\lambda_2}}
\newcommand{\lO}{{\lambda_1}}
\newcommand{\btau}{{\boldsymbol \tau}}

\newcommand{\mfs}{{\texttt M_{\texttt {FT}}}}
\newcommand{\mml}{{\texttt M_{\texttt {ML}}}}
\newcommand{\mrc}{{\texttt M_{\texttt {RC}}}}
\newcommand{\mbc}{{\texttt M_{\texttt {BC}}}}
\newcommand{\tmin}{\tau_{m}}

\DeclareMathOperator*{\argmin}{arg\,min}
\DeclareMathOperator*{\argmax}{arg\,max}

\def\hv{{\hat v}}

\renewcommand{\algorithmicrequire}{\textbf{Input:}}
\renewcommand{\algorithmicensure}{\textbf{Output:}}

\floatname{algorithm}{Protocol}

\makeatletter
\def\BState{\State\hskip-\ALG@thistlm}
\makeatother

\newfont{\mycrnotice}{ptmr8t at 7pt}
\newfont{\myconfname}{ptmri8t at 7pt}
\let\crnotice\mycrnotice%
\let\confname\myconfname%

\begin{document}

\title{Anonymity Properties of the Bitcoin P2P Network}

\numberofauthors{2}
\author{
\alignauthor Giulia Fanti \\
\affaddr{Coordinated Sciences Laboratory} \\
\affaddr{University of Illinois, Urbana-Champaign} \\
\affaddr{IL 61801} \\
\email{fanti@illinois.edu}
\alignauthor {Pramod Viswanath}\\
\affaddr{Coordinated Sciences Laboratory} \\
\affaddr{University of Illinois, Urbana-Champaign} \\
\affaddr{IL 61801} \\
\email{pramodv@illinois.edu}
}


\date{ }
\maketitle

\maketitle

\begin{abstract}
Bitcoin is a popular alternative to fiat money, widely used for its perceived anonymity properties.
However, recent attacks on Bitcoin's peer-to-peer (P2P) network demonstrated that its gossip-based flooding protocols, which are used to ensure global network consistency, may enable user deanonymization---the linkage of a user's IP address with her pseudonym in the Bitcoin network. 
In 2015, the Bitcoin community responded to these attacks by changing the network's flooding mechanism to a different protocol, known as diffusion. 
However, no systematic justification was provided for the change, and it is unclear if diffusion actually improves the system's anonymity.
In this paper, we model the Bitcoin networking stack and analyze its anonymity properties, both pre- and post-2015. 
In doing so, we consider new adversarial models and spreading mechanisms that have not been previously studied in the source-finding literature.
We theoretically prove that Bitcoin's networking protocols (both pre- and post-2015) offer poor anonymity properties on networks with a regular-tree topology.
We validate this claim in simulation on a 2015 snapshot of the real Bitcoin P2P network topology.
\end{abstract}
%
%
\category{G.2.2}{Graph Theory}{Network problems, Graph algorithms}
%
%
\keywords{Cryptocurrencies, Bitcoin, Peer-to-Peer Networks, Privacy} 

\section{Introduction}
The Bitcoin cryptocurrency has seen widespread adoption, due in part to its reputation as a privacy-preserving  financial system \cite{bitcoinprivacy,coinbase}.
In practice, though, Bitcoin exhibits a number of serious privacy vulnerabilities \cite{androulaki2013evaluating,meiklejohn2013fistful,reid2013analysis,ron2013quantitative,ober2013structure}.
Most of these vulnerabilities arise because of two key properties: (1) Bitcoin associates each user with a pseudonym, and (2) pseudonyms can be linked to financial transactions by way of a public transaction ledger, called the \emph{blockchain} \cite{bitcoin}. 
This means that if an attacker is able to associate a pseudonym with its human user, the attacker may learn the user's entire transaction history.
Such leakage represents a massive privacy violation, and would be deemed unacceptable in traditional banking systems.

In practice, there are several ways to link a user to her Bitcoin pseudonym.
The most commonly-studied methods analyze transaction patterns in the public blockchain, and link those patterns using side information \cite{ober2013structure,reid2013analysis}. 
In this paper, we are interested in a lower-layer vulnerability: the networking stack.
Like most cryptocurrencies, Bitcoin nodes communicate over a P2P network \cite{bitcoin}. 
The anonymity implications of this P2P network have been largely ignored until recently, when researchers demonstrated empirical deanonymization attacks that exploit the P2P network's management protocols \cite{biryukov,koshy2014analysis}. 
These findings are particularly troublesome because the Bitcoin P2P stack is used in a number of other cryptocurrencies (or \emph{altcoins}), some of which are designed with provable anonymity guarantees in mind \cite{zcash,sasson2014zerocash}.
Hence vulnerabilities in the Bitcoin P2P network may extend to a host of other cryptocurrencies as well.

We are interested in one key aspect of the Bitcoin P2P network:  the dissemination of transactions.
Whenever a user (Alice) generates a transaction (i.e., she sends bitcoins to another user, Bob), she first creates a ``transaction message" that contains her pseudonym, Bob's pseudonym, and the transaction amount.
Alice subsequently broadcasts this transaction message over the P2P network,
which enables other users to validate her transaction and incorporate it into the global blockchain.

The broadcast of transactions is critical to maintaining blockchain consistency.
Broadcasting proceeds by flooding transactions along links in the P2P network.
The actual flooding protocol has been the subject of some discussion, and is central to our paper. 
In particular, we are interested in theoretically quantifying the anonymity properties of existing flooding protocols used by the Bitcoin networking stack.

\vspace{0.1in}
\noindent \textbf{Anonymity in the Bitcoin  P2P network.}
Transaction broadcasting opens a new avenue for deanonymization attacks.
If an attacker can infer the IP address that initiated a transaction broadcast, 
then the attacker can also link the IP address to the associated user's Bitcoin pseudonym.
Since IP addresses can sometimes be linked to human identities (e.g., with the help of an ISP), 
this deanonymization vector is a powerful one.

In recent years, security researchers demonstrated precisely such deanonymization attacks, which exploit Bitcoin's transaction-flooding protocols.
These attacks rely on a ``supernode" that connects to active Bitcoin nodes and listens to the transaction traffic relayed by honest nodes \cite{koshy2014analysis,biryukov,biryukov2015bitcoin}.
Using this technique, researchers were able to link Bitcoin users' pseudonyms to their IP addresses with an accuracy of up to 30\% \cite{biryukov}.

In 2015, the Bitcoin community responded to these attacks by changing its flooding protocols from a gossip-style  protocol known as \emph{trickle spreading} to a \emph{diffusion spreading} protocol that spreads content with independent exponential delays \cite{bitcoindTrickleDiffusion}. 
We define these protocols precisely in Section \ref{sec:model}.
However, no systematic motivation was provided for this shift. 
Indeed, it is unclear whether the change actually defends against the deanonymization attacks in \cite{biryukov,koshy2014analysis}. 


\subsection{Problem and Contributions}
Our goal is to analyze the anonymity properties of the Bitcoin P2P network.
The main point of our paper is to show that the Bitcoin network has poor anonymity properties, 
and the community's shift from trickle spreading (pre-2015) to diffusion spreading (post-2015) did not help the situation. 
The optimal (maximum-likelihood) source-identification algorithms change between protocols;
identifying such algorithms and quantifying their performance is the primary focus of this work.
We find that despite having different maximum-likelihood estimators, trickle and diffusion exhibit roughly the same, poor anonymity properties.
Our specific contributions are threefold:

\vspace{0.07in}
\noindent \textbf{(1) Modeling.} We model the Bitcoin P2P network and an `eavesdropper adversary', whose capabilities reflect recent practical attacks in \cite{biryukov,koshy2014analysis}. 
This task is complicated by the fact that most Bitcoin network protocols are not explicitly documented. 
Modeling the system therefore requires parsing a combination of documentation, papers, and code. 
Several of the resulting models are new to the rumor source analysis literature.

\vspace{0.07in}
\noindent \textbf{(2) Analysis of Trickle (Pre-2015).} We analyze the probability of deanonymization by an eavesdropper adversary under trickle propagation, which was used until 2015. 
Our analysis is conducted over a regular tree-structured network.
Although the Bitcoin network topology is not a regular tree, we will see in Section \ref{sec:model} that regular trees are a reasonable first-order model.
We consider suboptimal, graph-independent estimators (e.g., the first-timestamp estimator), as well as maximum-likelihood estimators; both are defined precisely in Section \ref{sec:model}. 
Our analysis suggests that although the first-timestamp estimator performs poorly on high-degree trees, maximum-likelihood estimators can achieve high probabilities of detection for trees of any degree $d$ (Table \ref{tab:results}).

\vspace{0.07in}
\noindent \textbf{(3) Analysis of Diffusion (Post-2015).} 
We conduct a similar analysis of diffusion spreading, which was adopted in 2015 as a fix for the anonymity weaknesses observed under trickle propagation \cite{biryukov,koshy2014analysis}. 
Table \ref{tab:results} summarizes a subset of our results, which characterize the probability of detection  asymptotically in tree degree $d$. 
We wish to highlight the fact that trickle and diffusion exhibit similar anonymity behavior.
We revisit this table  more carefully in Section \ref{sec:experiments}, where we also empirically validate our findings on a snapshot of the real Bitcoin network from 2015. 

\begin{table}[t]
\centering
\caption{Summary of  probability of detection results on regular trees. Asymptotics are in the tree degree $d$. Notice that results for trickle and diffusion are similar. }
\label{tab:results}
\begin{tabular}{|c|c|c|}
\hline
                                                           & \begin{tabular}[c]{@{}c@{}}{\bf Trickle} \\ {\bf (pre-2015)} \end{tabular}&\begin{tabular}[c]{@{}c@{}} {\bf Diffusion} \\ {\bf (post-2015)}\end{tabular} \\ \hline
\begin{tabular}[c]{@{}c@{}}{\bf First-}\\ {\bf Timestamp}\end{tabular} &    \begin{tabular}{@{}c@{}} $\frac{\log(d)}{d \log(2)} + o\left( \frac{\log d}{d}\right)$ \\ (Eq. \ref{eq:trickle1d})  \end{tabular}  &     \begin{tabular}{@{}c@{}} $\frac{\log(d-1)}{(d-2)}$ \\ (Thm. \ref{thm:fs_diffusion})     \end{tabular}   \\ \hline
\begin{tabular}[c]{@{}c@{}}{\bf Maximum-}\\ {\bf Likelihood}\end{tabular}                                                 &  $\Theta (1)$ ~ (Thm. \ref{thm:ml_trickle})        &     $\Theta (1)$ ~ (Thm \ref{thm:rc_diffusion})       \\ \hline
\end{tabular}
\end{table}

\vspace{0.1in}
\noindent \textbf{Paper Structure.}
We begin by modeling Bitcoin's P2P networking stack and the adversaries of interest in Section \ref{sec:model}.
We then analyze the performance of trickle propagation in Section \ref{sec:trickle}.
In Section \ref{sec:diffusion}, we analyze the performance of diffusion.
We compare these results side-by-side in Section \ref{sec:experiments}, which also includes empirical trials on a snapshot of the real Bitcoin network.
We discuss the relation between our results and prior work in Section \ref{sec:related}.
Section \ref{sec:discussion} concludes by discussing the practical implications of these results and preliminary ideas on how to solve the problem.

\section{Model and Problem Statement}
\label{sec:model}

To characterize the anonymity of Bitcoin's P2P network, we need to model three key aspects of the system:
the network topology, the spreading protocol, and the adversary's capabilities. 

\subsection{Network Model}
The Bitcoin P2P network contains two classes of nodes: servers and clients. 
Clients are nodes that do not accept incoming TCP connections (e.g., nodes behind NAT), whereas
servers do accept incoming connections.
Clients and servers have different networking protocols and anonymity concerns. 
For instance, clients do not relay transactions.
We focus in this work on servers. 

We model the P2P network of servers as a graph $G(V,E)$, where $V$ is the set of all server nodes and $E$ is the set of edges, or connections, between them.
In practice, each server node is represented by a (IP address, port) tuple.
Currently, there are about 5,000 active Bitcoin servers, and this number generally remains stable over the timescale of a single transaction broadcast  \cite{bitnodes}.
Each server is allowed to establish up to eight outgoing connections 
to active Bitcoin nodes and maintain up to 125 total active connections \cite{biryukov,bitcoind}. 
For a connection between Alice and Bob, an \emph{outgoing connection} (from Alice's perspective) is one that is initiated by Alice, whereas an \emph{incoming connection} is one initiated by Bob. 
However, these TCP connections are bidirectional once established. 
Most of a server's incoming connections are from Bitcoin clients, so they are irrelevant to our analysis. 
However, we will see momentarily that adversaries can use the asymmetry between incoming and outgoing connections to monitor the network for deanonymization attacks.

The resulting sparse random graph between servers can be modeled approximately as a 16-regular graph; in practice, the average degree is closer to 8 due to nonhomogeneities across nodes \cite{coinscope}. 
Critically, the graph is locally tree-like and (approximately) regular.
For this reason, \textbf{regular trees} are a natural class of graphs to study.
In our theoretical analysis, we model $G$ as a $d$-regular tree.
We validate this choice by running simulations on a snapshot of the true Bitcoin network \cite{coinscope} (Section \ref{sec:experiments}).


\subsection{Spreading Protocols}
Recall that every time a transaction (or a block) is completed, it is broadcast over the network. 
In this work, we analyze the spread of a single message originating from source node $\vs \in V$. 
Without loss of generality, we will label $\vs$ as node `0' whenever we are iterating over nodes. 

The broadcasting protocol for  messages has been under discussion in the Bitcoin community recently. 
Until 2015, the Bitcoin network used a gossip-like \emph{trickle} broadcasting protocol.
However, in the wake of various anonymity attacks \cite{biryukov,biryukov2015bitcoin,koshy2014analysis}, the reference Bitcoin implementation changed its networking stack to use a different broadcasting protocol known as \emph{diffusion} \cite{bitcoindTrickleDiffusion}.
In this paper, we evaluate both protocols and compare their performance.

\textbf{Trickle spreading} is a gossip-based flooding protocol. 
Each message source or relay randomly orders its neighbors who have not yet seen the message; we call these \emph{uninfected} neighbors. 
It then transmits the message to its neighbors according to the ordering, with a constant delay of 200 ms between transmissions \cite{biryukov}.
We model this spreading protocol by assuming a discrete-time system; each source or relay randomly orders its uninfected neighbors and transmits the message to one neighbor per subsequent time step. 
We assume a node begins relaying a message in the first timestep after it receives the message.

In \textbf{diffusion spreading}, each source or relay node transmits the message to each of its uninfected neighbors with an independent, exponential delay of rate $\lambda$.
We assume a continuous-time system, in which  a node starts the exponential clocks as soon as it receives (or creates) a message.

For both protocols, we let $X_v$ denote the timestamp at which honest node $v\in V$ receives a given message. 
Note that server nodes cannot be infected more than once.
We  assume the message originates at time $t=0$, so $X_{\vs} =X_0=0$.
Moreover, we let $G_t(V_t, E_t)$ denote the \emph{infected subgraph} of $G$ at time $t$, or the subgraph of nodes who have received the message (but not necessarily reported it to the adversary) by time $t$.

\subsection{Adversarial Model}
We consider an adversary whose goal is to link a message with the (IP address, port) that originated it. 
In our setup, this translates to identifying the source node $\vs \in V$.

To this end, we introduce an  \textbf{eavesdropper adversary}, whose capabilities are modeled on the practical deanonymization attacks in \cite{biryukov,koshy2014analysis}.
These attacks are cheap, scalable, and simple, so they represent realistic threats to the network.
We begin by describing the attacks at a practical level, and then extract models for the actual analysis.

\begin{figure}[t]
    \centering
  \includegraphics[width=.3\textwidth]{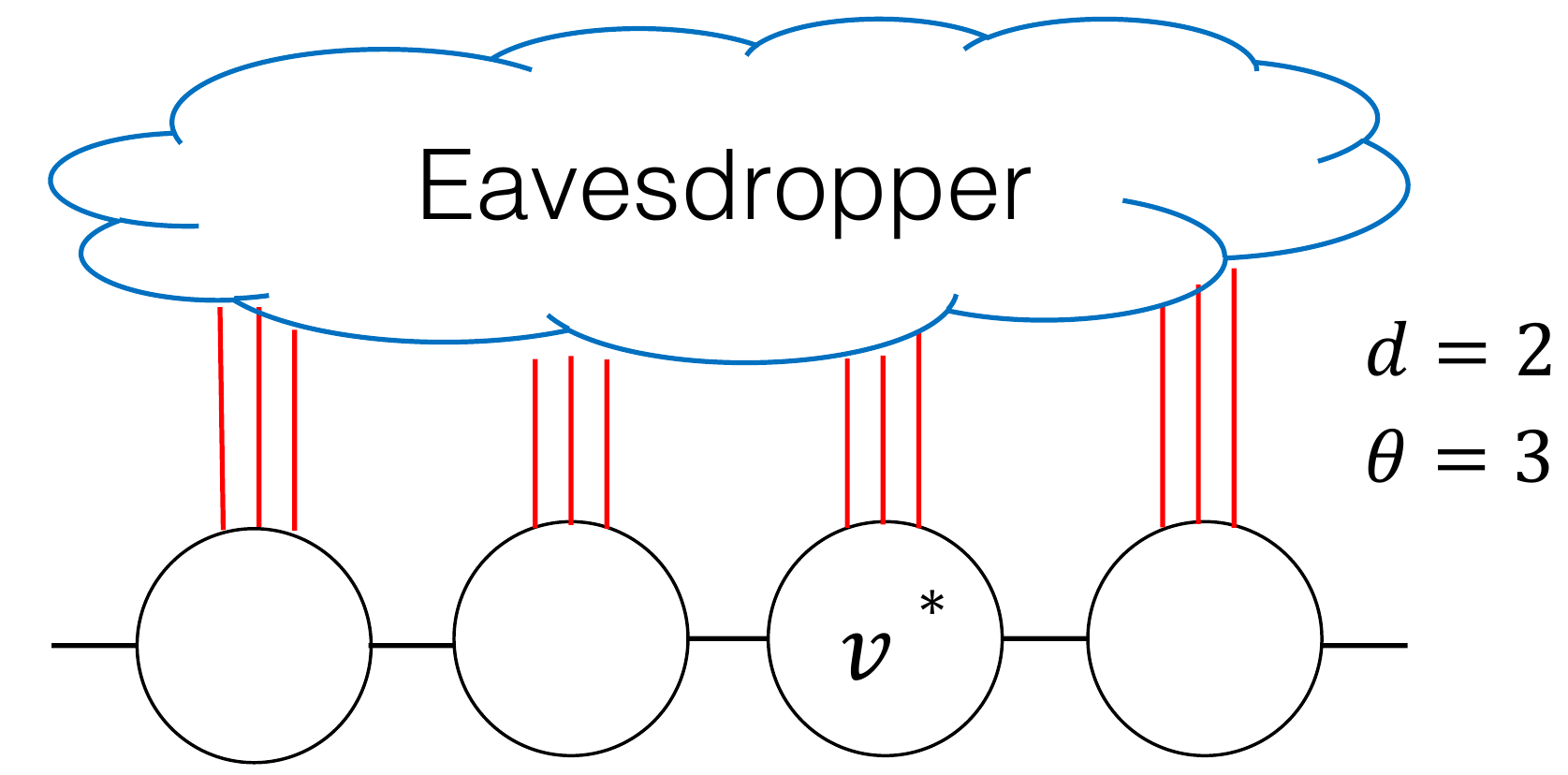}
  \caption{The eavesdropper adversary establishes $\theta$ links (shown in red) to each server. 
  Honest servers are connected in a $d$-regular tree topology (edges shown in black).}
  \label{fig:eavesdropper}
\end{figure}

The attacks in \cite{biryukov,koshy2014analysis} use a supernode that connects to most of the servers in the Bitcoin network. 
The supernode can make multiple connections to each honest server, with each connection coming from a different (IP address, port). 
Hence, the honest server does not realize that the supernode's connections are all from the same entity.
The supernode can compromise arbitrarily many of a server's unused connections, up to the hard limit of 125 total connections.
We model this setup  by assuming that the eavesdropper adversary makes a fixed number $\theta$ of connections to each server, where $\theta \geq 1$. 
We do not include these adversarial connections in the original server graph $G$, so $G$ remains a $d$-regular graph.
This setup is illustrated in Figure \ref{fig:eavesdropper}.

Once the supernode in \cite{biryukov,koshy2014analysis} is established, it simply listens to all relayed messages on the network, without relaying or transmitting any content---hence the name `eavesdropper adversary.'
Over time, due to peer address propagation protocols (which were not discussed here), the adversary learns the network structure between servers. 
Therefore, we assume that  $G(V,E)$ is known to the eavesdropper adversary.

The supernode in \cite{biryukov,koshy2014analysis} also observes the timestamps at which messages are relayed from each honest server.
Since the adversary maintains multiple active connections to each server, it receives the message multiple times from each server. 
For ease of analysis, we assume that the eavesdropper adversary only stores the \emph{first} such timestamp. 
We also assume that the adversary observes all timestamps relative to time $t=0$, 
i.e., it knows when the message started spreading. 
Given this, we let $\tau_v$ denote the time at which the adversary first observes the message from node $v\in V$.  
 We let $\boldsymbol \tau$ denote the set of all observed timestamps.

\vspace{0.1in}
\noindent \textbf{Relation to existing adversarial models.}
We wish to briefly highlight the difference between the eavesdropper adversary and two commonly-studied adversarial models. 
The \textbf{snapshot adversary} is the most common adversary in this space \cite{SZ11a,SZ11b,SZ12,FC12,LMOZ13}.
A snapshot adversary observes the set of infected nodes at a single time $T$; in our notation, the adversary learns the set $\{v\in V : X_v \leq T\}$ (no timestamps), along with graph $G$. 
The eavesdropper adversary differs in that it eventually observes a noisy timestamp $\tau_v$ for \emph{every} node, regardless of when the node is infected. 

Another common adversarial model is the \textbf{spy-based adversary}, which observes exact timestamps for a corrupted set of nodes that does not include the source \cite{PTV12,ZY13}.
In our notation, for a set of spies $S \subseteq V$, the spy-based adversary observes $\{(s, X_s) :~s\in S \}$.
This is different from the eavesdropper adversary because the eavesdropper only observes delayed timestamps, and it does so for \emph{all} nodes, including the source.
Precise analysis of the spy-based adversary has not appeared in the literature.

Neither of these prior adversarial models adequately captures the recent supernode-based deanonymization attacks on the Bitcoin network \cite{biryukov,koshy2014analysis}.
Indeed, we shall see that the eavesdropper adversary requires different analytical techniques from traditional adversarial models.
On the other hand, some of the techniques we will use to analyze the eavesdropper adversary can be applied to traditional adversaries (Sec. \ref{sec:experiments}).

\vspace{0.1in}
\noindent \textbf{Source Estimation.}
The adversary's goal is as follows: given the observed noisy timestamps $\boldsymbol \tau$ (up to estimation time $t$) and the server graph $G$, find an estimator $\texttt M(\boldsymbol \tau, G)$ that correctly identifies the true source.
Our metric of success for the adversary is \textbf{probability of detection}. 
Given an estimator $\texttt M$, the adversary's probability of detection is $\prob(\texttt M(\boldsymbol \tau, G)=v^*)$.
The probability is taken over the random  spreading realization (captured by $\boldsymbol \tau$).

In \cite{biryukov,koshy2014analysis}, the adversary's estimator is a variant of the so-called \textbf{first-timestamp estimator}.
The first-timestamp estimator $\mfs(\boldsymbol \tau, G)$ outputs the first node (prior to estimation time $t$) to report the message to the adversary: 
$$
\mfs(\boldsymbol \tau, G) = \argmin_{v \in V_t} \tau_v.
$$
The first-timestamp estimator is popular because of its simplicity: it requires no knowledge of the graph, and it is computationally easy to implement.
Despite its simplicity, the first-timestamp estimator achieves high accuracy rates in practice \cite{biryukov,koshy2014analysis}.
We begin by analyzing this estimator for both trickle and diffusion propagation.

In principle, the adversary could implement more complicated estimators that utilize the underlying graph structure.
We are particularly interested in the \textbf{maximum-likelihood (ML) estimator}, which maximizes the adversary's probability of detection:
$$
\mml(\boldsymbol \tau, G) = \argmax_{v \in V} \prob(\boldsymbol \tau |G, \vs=v).
$$
The ML estimator depends on the time of estimation $t$ to the extent that $\btau$ only contains timestamps up to time $t$.
Unlike the first-timestamp estimator, the ML estimator differs across spreading protocols, depends on the graph, and may be computationally intractable. 
We nonetheless bound its performance for both trickle and diffusion spreading over regular trees.

\subsection{Problem Statement}

Our primary goal is to understand whether the Bitcoin community's move from trickle spreading to diffusion actually improved the system's anonymity guarantees. 
As such, our main problem is to characterize the maximum-likelihood (ML) probability of detection of the eavesdropper adversary for both trickle and diffusion processes on $d$-regular trees, as a function of degree $d$, number of corrupted connections $\theta$, and detection time $t$. 
We meet this goal by computing lower bounds derived from the analysis of  suboptimal estimators (e.g., first-timestamp estimator and centrality-based estimators), and upper bounds derived from fundamental limits on detection.

\section{Analysis of Trickle (Pre-2015)}
\label{sec:trickle}
We begin by analyzing the probability of detection of trickle spreading. 
We first consider the first-timestamp estimator, followed by the ML estimator.

\subsection{First-Timestamp Estimator}
The analysis of trickle propagation is complicated by its combinatorial, time-dependent nature. 
As such, we begin by lower-bounding the first-timestamp estimator's probability of detection.
We do so by computing the probability that the true source reports the message strictly before any other node.
Let 
$$
\tau_{m} \triangleq \min(\tau_1, \tau_2, \ldots)
$$
denote the minimum observed timestamp among nodes that are \emph{not} the source. 
Then we compute $\prob(\tau_0 < \tau_m)$, i.e., the probability that the true source reports the message to the adversary strictly before any of the other nodes.
This event (which causes the source to be detected with probability 1) does not include cases  where the true source is one of $k$ nodes ($k>1$) that report the message to the adversary simultaneously, and before any other node in the system.  
Since $\prob(\tau_0 < \tau_m)$ does not account for such `simultaneous reporting' events,  it is a lower bound.
Nonetheless, for large $d$, the `simultaneous reporting' event is rare, so our lower bound is close to the empirical probability of detection of the first-timestamp estimator.
\begin{theorem} \label{thm:fs_trickle}
Consider a message that propagates according to trickle spreading over a $d$-regular tree of honest servers, where each node additionally has $\theta$ connections to an eavesdropping adversary.
The first-timestamp estimator's probability of detection at time $t =\infty$ satisfies
\begin{align}
\prob(\mfs(\boldsymbol \tau, G) = \vs) &\geq& \frac{\theta}{d\log 2} \left[\text{Ei}(2^d \log \rho) -\text{Ei}\left ( \log \rho \right ) \right ]
\label{eq:pd_trickle_ei}
 \end{align}
 where $\rho = \frac{d-1}{d-1+\theta}$, Ei$(\cdot)$ denotes the exponential integral, defined as
 \[
 \text{Ei}(x) \triangleq -\int_{-x}^\infty \frac{e^{-t}dt}{t},
 \] 
and all logarithms are natural logs.
\end{theorem}
(Proof in Section \ref{proof:fs_trickle})

We prove this bound by conditioning on the time at which the source reports to the adversary, and computing the conditional probability that all other nodes report later.
The proof then becomes a combinatorial counting problem.

\vspace{0.1in}
\noindent \textbf{Implications.}
We can approximate the asymptotic behavior of equation \eqref{eq:pd_trickle_ei} for large $d$ 
by using the exponential integral's Taylor expansion. 
First, we note that when $d$ is large, $\text{Ei}(2^d \log \rho)\approx 0$, so we have 
\begin{align}
&\frac{\theta}{d\log 2} \left[\text{Ei}(2^d \log \rho) -\text{Ei}\left ( \log\rho \right ) \right ] \nonumber \\
									  \approx& \small{\frac{\theta }{d\log 2} \left ( -\gamma - \log \left | \log \rho \right | - \sum_{\nu = 1}^{\infty} \frac{ (\log \rho)^\nu}{\nu \cdot \nu!} \right )} \label{eq:approx_ei} \\
									  \approx & \frac{\theta }{d\log 2} \left ( -\gamma - \log  \log \left ( 1 +\frac{\theta}{d} \right )+ \log(1+\frac{\theta}{d}) \right.  \nonumber \\
									  & \left.  \quad -\frac{\log^2(1+\frac{\theta}{d})}{4} + \ldots \right ) \label{eq:bigd}\\
									  \approx & \frac{\theta }{d\log 2} \left ( -\gamma - \log \frac{\theta}{d} + \frac{\theta}{d} - \frac{\theta^2}{4d^2} + \ldots \right ) \label{eq:smalltheta}
\end{align}
where $\gamma \approx 0.577$ is the Euler-Mascheroni constant \cite{weisstein2002euler},  and \eqref{eq:approx_ei} comes from substituting the exponential integral by its Taylor expansion for real arguments \cite{bender1999advanced}. 
Line \eqref{eq:bigd} holds because as $d\rightarrow \infty$, $\rho \approx \frac{1}{1+\theta/d}$, 
and \eqref{eq:smalltheta} holds because as $d \rightarrow \infty$,  $\log(1+\frac{\theta}{d})\approx \frac{\theta}{d}$.

In particular, for the special case of $\theta = 1$ where the adversary establishes only one connection per server, line \eqref{eq:smalltheta} simplifies to
\begin{eqnarray}
\prob(\mfs(\boldsymbol \tau, G)) \approx   \frac{\log d}{d \cdot \log 2} + o \left ( \frac{\log d}{d} \right ).
\label{eq:trickle1d}
\end{eqnarray}
This suggests that the first-timestamp estimator has a  probability of detection that decays to zero asymptotically as $\log(d)/d$.
Intuitively,  the probability of detection should decay to zero, because the higher the degree of the tree, the higher the likelihood that a node \emph{other} than the source reports to the adversary before the source does. 
Nonetheless, \eqref{eq:trickle1d} is only a lower bound on the first-timestamp's probability of detection, so we wish to understand how tight the bound is.

\begin{figure}[t]
    \centering
  \includegraphics[width=.4\textwidth]{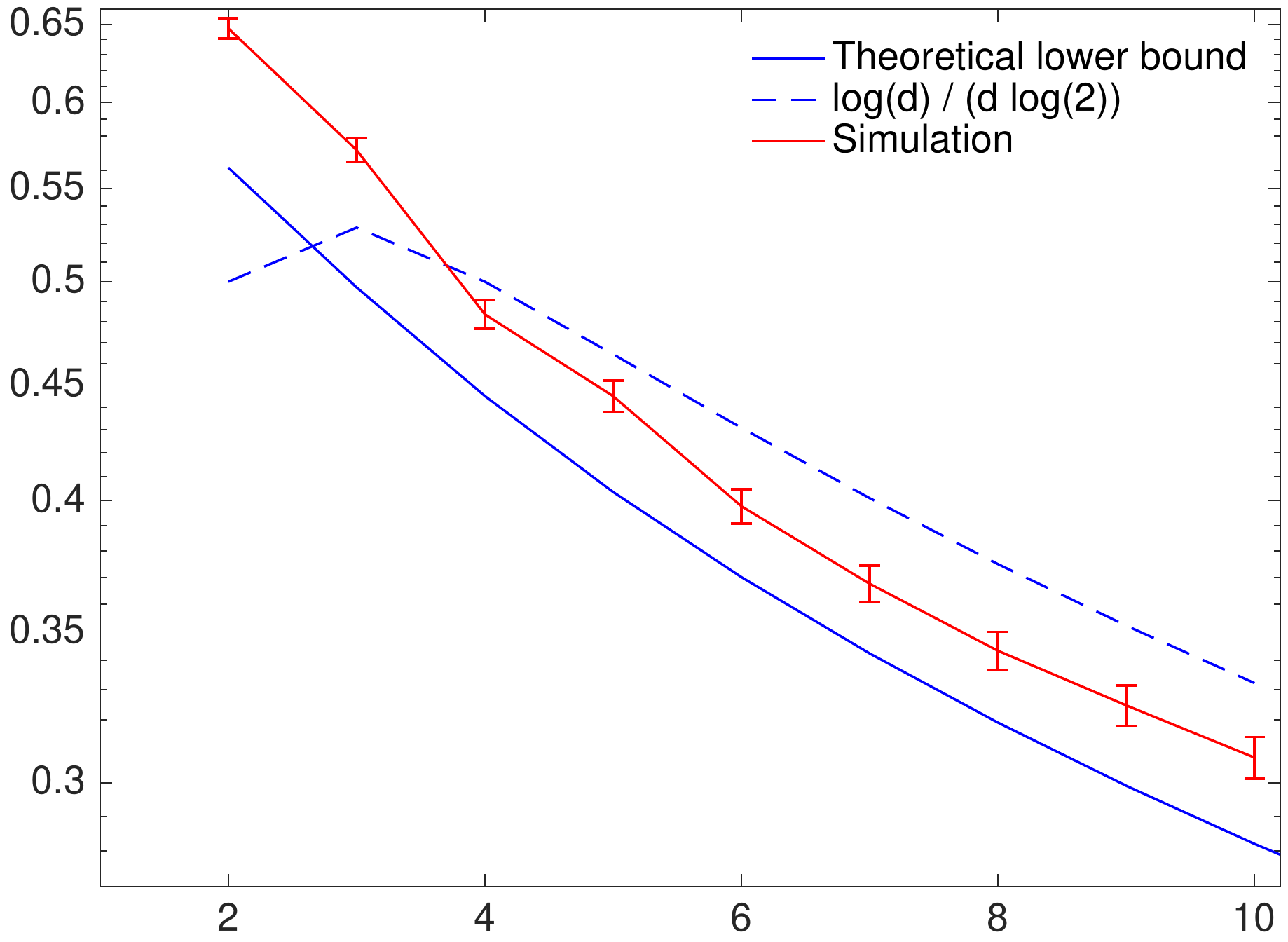}
  \put(-215,30){\rotatebox{90}{Probability of Detection}}
  \put(-120,-10){Tree degree, $d$}
  \caption{First-timestamp estimator accuracy on $d$-regular trees when the adversary has one link per server ($\theta = 1$).}
  \label{fig:fs_trickle_1d}
\end{figure}

\vspace{0.1in}
\noindent \textbf{Simulation Results.}
To evaluate the lower bound in Theorem \ref{thm:fs_trickle} and the approximation in \eqref{eq:trickle1d}, we simulate the first-timestamp estimator on regular trees. 
Figure \ref{fig:fs_trickle_1d} illustrates the simulation results for $\theta = 1$ compared to the approximation in \eqref{eq:trickle1d}. 
Each data point is averaged over 5,000 trials.
In practice, the lower bound appears to be tight, especially as $d$ grows. 

Figure \ref{fig:fs_trickle_1d} and approximation \eqref{eq:trickle1d} suggest a natural solution to the Bitcoin network's anonymity problems: increase the degree of each node to reduce the adversary's probability of detection. 
However, we shall see in the next section that stronger estimators (e.g., the ML estimator) may achieve high probabilities of detection, even for large $d$.
Therefore, the trickle protocol is unlikely to adequately protect users' anonymity.
 
\subsection{Maximum-Likelihood Estimator}
In this section, we identify an ML source estimator for finite detection time $t$, which uses the graph structure and timestamps to infer the source. 
We analyze its limiting probability of detection as $t\rightarrow \infty$, and show that it achieves a probability of detection lower-bounded by 1/2, regardless of the degree of the tree. 
This highlights the weakness of trickle propagation when the adversary knows the graph.
We begin with a discussion of the ML estimator.

As the message spreads, it reaches some nodes before others. 
At any time $t$, if one knew the ground truth timestamps (i.e., the $X_v$'s), one could arrange the nodes of the infected subgraph $G_t$ in the order they received the message.
We call such an arrangement an \emph{ordering} of the nodes.
Since trickle propagation is a discrete-time system, multiple nodes may receive the message simultaneously, in which case they are lumped together in the ordering.
Of course, the true ordering is not observed by the adversary, but the observed timestamps (i.e., $\boldsymbol \tau$) restrict the set of possible orderings.
A \emph{feasible ordering} is an ordering that respects the rules of trickle propagation over graph $G$, as well as the observed timestamps $\boldsymbol \tau$.
In this subsection only, we will abuse notation by using $\boldsymbol \tau$ to refer to \emph{all} timestamps observed by the adversary, not just the first timestamp from each server. 
So if the adversary has $\theta$ connections to each server, $\btau$ would include $\theta$ timestamps per honest server.

We propose an estimator called \textbf{timestamp rumor centrality}, which counts the number of feasible orderings originating from each candidate source.
The candidate with the most feasible orderings is chosen as the estimator output. 
This estimator is similar to rumor centrality, an estimator devised for snapshot adversaries in \cite{SZ10}.
However, the presence of timestamps and the lack of knowledge of the infected subgraph complicates matters.
We first motivate timestamp rumor centrality, then show that it is the ML source estimator for trickle spreading under an eavesdropper adversary. 

\begin{propo}\label{lem:ordering}
Consider a trickle process over a $d$-regular graph, where each node has $\theta$ connections to the eavesdropper adversary.
Any feasible orderings $o_1$ and $o_2$ with respect to observed timestamps $\boldsymbol \tau$ and graph $G$  have the same likelihood.
\end{propo}
(Proof in Section \ref{proof:ordering})

This claim can be proved with an inductive argument, which shows that at any given time, the number of nodes with a given \emph{uninfected degree}, or number of uninfected neighbors, is deterministic---i.e., it does not depend on the underlying ordering. 
Moreover, the likelihood of a given ordering is strictly a function of the nodes' uninfected degrees at each time step, so all feasible orderings have the same likelihood. 

Proposition \ref{lem:ordering} implies that at any fixed time, the likelihood of observing $\btau$ given a candidate source is proportional to the number of feasible orderings originating from that candidate source. 
Therefore, an ML estimator, which we call timestamp rumor centrality, counts the number of feasible orderings at finite estimation time $t$.

Timestamp rumor centrality is a message-passing algorithm that proceeds as follows: for each candidate source, recursively determine the set of feasible times when each node could have been infected,
given the observed timestamps. 
This is achieved by passing a set of ``feasible times of receipt" from the candidate source to the leaves of the largest feasible subtree rooted at the candidate source.
In each step, nodes prune any times of receipt that conflict with their observed timestamps. 
Next, given each node's set of feasible receipt times, they count the number of feasible orderings that obey the rules of trickle propagation.
This is achieved by passing sets of partial orderings from the leaves back to the candidate source, and pruning out infeasible orderings.

In practice, we do not pass the entire partial ordering; we can instead store the \emph{number} of distinct partial orderings for the subtree rooted at each child, indexed by each feasible time of receipt for the child (there are $O(d)$ feasible times of receipt).
This reduces the overall computational complexity to $O((2d)^d|V|)$: each node passes a message of size $O(d)$ to its parent, and each parent takes a Cartesian product of its children's messages. 
The whole procedure is run for each candidate source, of which there are $O(2^d)$, since the true source has a timestamp of at most $d+1$.
Due to the additional notation needed to describe timestamp rumor centrality in detail, we defer a full-fledged pseudocode description to Appendix \ref{app:algorithms} (Figure \ref{algo:timestamp}). 

In \cite{SZ12}, precise analysis of standard rumor centrality was possible because rumor centrality can be reduced to a simple counting problem.
Such an analysis is more challenging for timestamp rumor centrality, because timestamps prevent us from using the same counting argument. 
However, we identify a suboptimal, simplified version of timestamp rumor centrality that approaches optimal probabilities of detection as $t$ grows. 
We call this estimator \textbf{ball centrality}.

\begin{figure}[t]
    \centering
  \includegraphics[width=.45\textwidth]{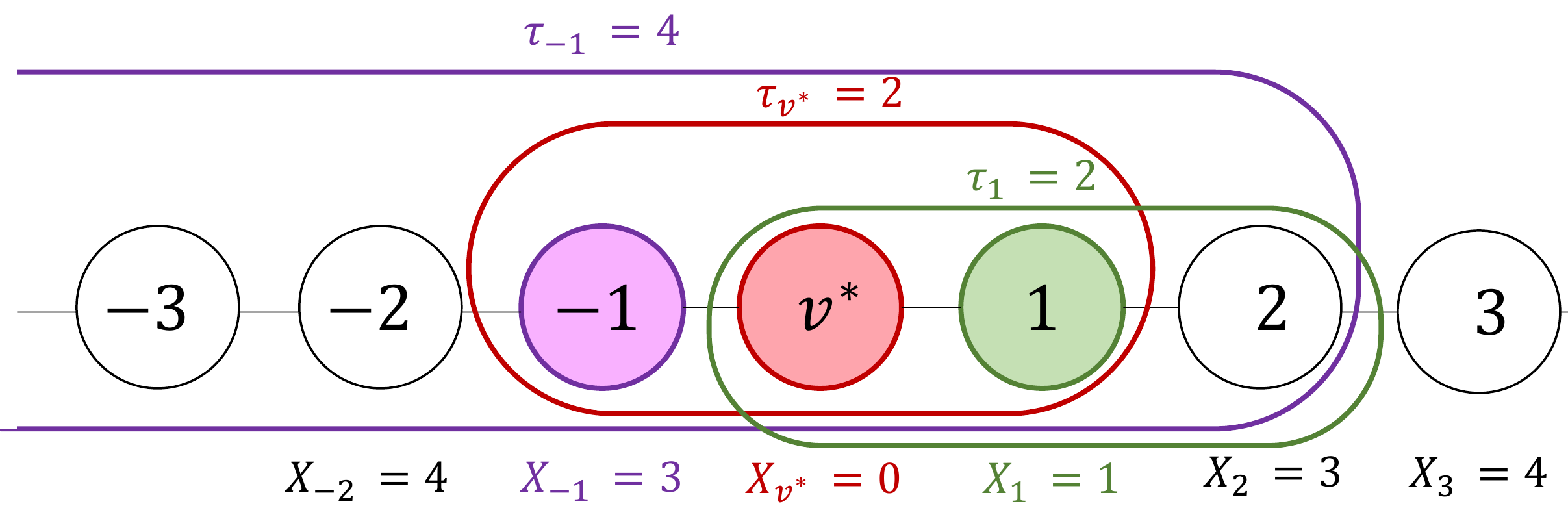}
  \caption{Example of ball centrality on a line with one link to the adversary per server (these links are not shown). The estimator is run at time $t=4$.}
  \label{fig:ball}
\end{figure}

Unlike timestamp rumor centrality, ball centrality simply checks whether a candidate source $v$ could have generated each of the observed timestamps, independently.
We explain this in the context of an example.
Figure \ref{fig:ball} contains a sample spread on a line graph, where the adversary has one connection per server (not shown). 
Therefore, $d=2$ and $\theta = 1$.
The ground truth infection time is written as $X_v$ below each node, and the observed timestamps are written above the node. 
In this figure, the estimator is run at time $t=4$, so the adversary only sees three timestamps.
For each observed timestamp $\tau_v$,
the estimator creates a ball of radius $\tau_v-1$, centered at $v$.
For example, in our figure, the green node (node 1) has $\tau_1=2$. 
Therefore, the adversary would make a ball of radius 1 centered at node 1; 
this ball is depicted by the green bubble in our figure.
The ball represents the set of nodes that are close enough to node 1 to feasibly report to the adversary from node 1 at time $\tau_1=2$.
After constructing an analogous ball for every observed timestamp in $\btau$, the protocol outputs a source selected uniformly from the intersection of these balls.
In our example, there are exactly two nodes in this intersection.
We outline ball centrality precisely in Protocol \ref{algo:ball}.

\begin{algorithm}[b]
\caption{{\sc Ball Centrality}. Returns a source estimate whose location is consistent with timestamps $\btau$ on tree $G$. $h(v,w)$ denotes the hop distance between $v$ and $w$.}
\label{algo:ball}
\begin{algorithmic}[1]
\renewcommand{\algorithmicrequire}{\textbf{Input:}}
\renewcommand{\algorithmicensure}{\textbf{Output:}}
\Require Timestamps $\btau$, graph $G(V,E)$
\Ensure Source estimate $\hat v\in V$
\State $W\gets V$
\For {$v \in V$}
\Comment Find the intersection of feasible balls
	\State $W \gets W \cap \{w\in V: h(w,v)\leq \tau_w-1\}$
\EndFor
\State $\hat v\sim \text{Unif}(W)$
\end{algorithmic}
\end{algorithm}

Although ball centrality is not ML for a fixed time $t$, the following theorem lower bounds the ML probability of detection by analyzing ball centrality and showing that its probability of detection approaches a fundamental upper bound exponentially fast in detection time $t$.

\begin{theorem} \label{thm:ml_trickle}
Consider a trickle spreading process over a $d$-regular graph of honest servers. In addition, each server has $\theta$ independent connections to an eavesdropper adversary. 
The ML probability of detection at time $t$ satisfies the following two expressions:
\begin{eqnarray}
(1)& \qquad \prob(\mml(\boldsymbol \tau, G) = \vs) \leq 1 - \frac{d}{2(\theta + d)}   \label{eq:ub} \\
(2)& \qquad \prob(\mml(\boldsymbol \tau, G) = \vs) \geq 1 - \frac{d}{2(\theta +d)} - \left ( \frac{d}{\theta + d}\right ) ^t   \label{eq:lb}
\end{eqnarray}
\end{theorem}
(Proof in Section \ref{proof:ml_trickle})

To prove the upper bound, the key idea is as follows:  in the first time step, if a source infects an honest neighbor instead of the adversary, then the true source and the honest neighbor become indistinguishable to the adversary. 
This is because the spreading processes from each node after that first time step become identically distributed. 
This gives an upper bound on the probability of detection.
For the lower bound, the key idea is that if we wait long enough, the ball estimator will eventually return an intersection of balls containing at most two nodes: the source and the first honest neighbor to whom it passed the message. 
We can characterize the probability of this happening as a function of time,
which lower bounds the ML probability of detection.
  
We make a few additional remarks about Theorem \ref{thm:ml_trickle}:

\vspace{0.07in}
\noindent (1) The right-hand side of equation  \eqref{eq:ub} is always greater than $\frac{1}{2}$, regardless of $\theta$. 
As such, increasing the degree of the graph would not significantly reduce the probability of detection---the adversary can still identify the source with probability at least $\frac{1}{2}$, given enough time.

\begin{figure}[t]
    \centering
  \includegraphics[width=.4\textwidth]{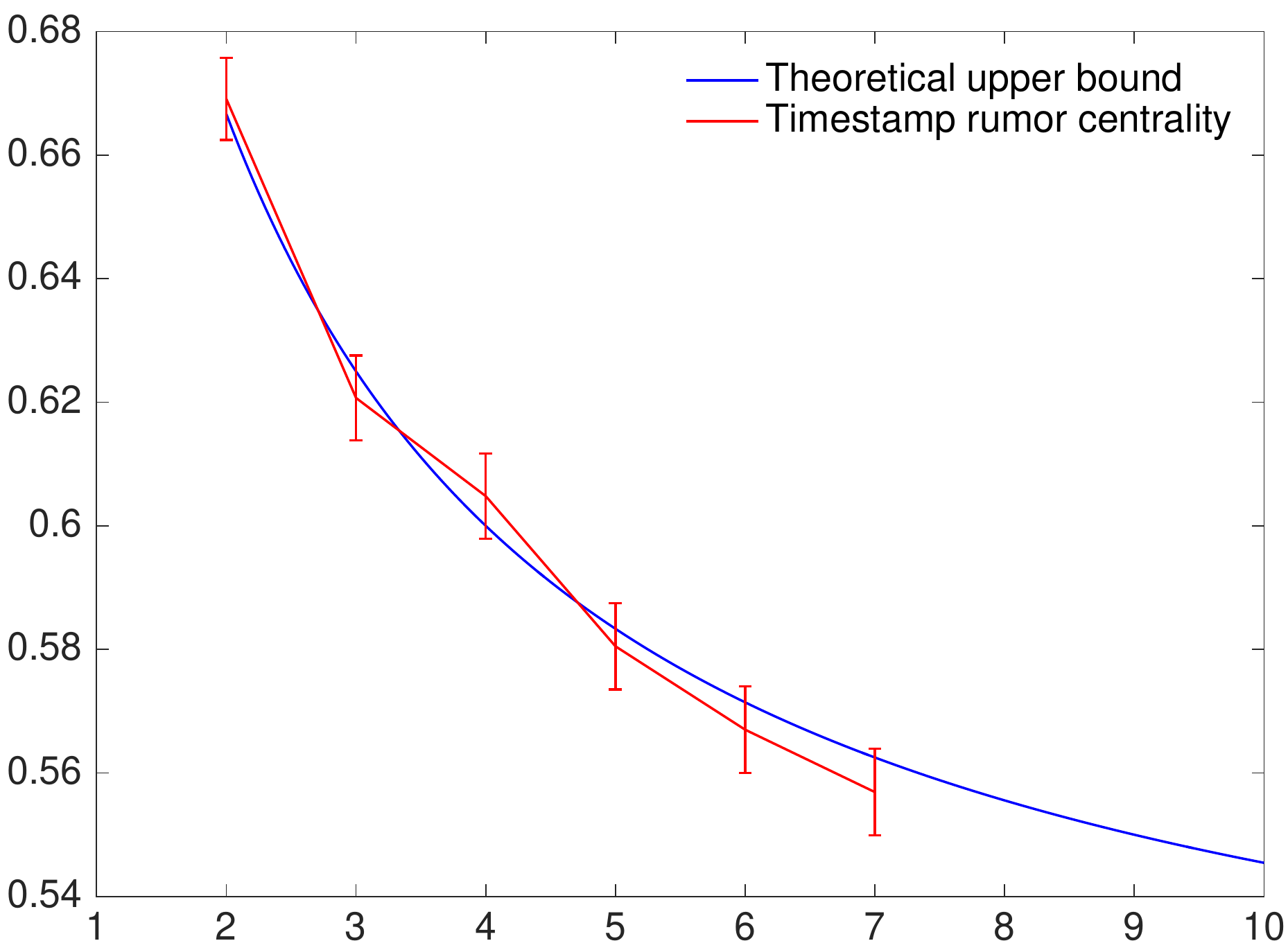}
  \put(-215,30){\rotatebox{90}{Probability of Detection}}
  \put(-120,-10){Tree degree, $d$}
  \caption{Timestamp rumor centrality (ML estimator) accuracy on $d$-regular trees when the adversary has one link per server ($\theta = 1$). The estimator is run at time $t=d+1$.}
  \label{fig:ml_trickle_1d}
\end{figure}

\vspace{0.07in}
\noindent (2) The ML probability of detection approaches its upper bound exponentially fast in time $t$. 
This suggests that the adversary can, in practice, achieve high probabilities of detection at small times $t$.
For example, Figure \ref{fig:ml_trickle_1d} shows the probability of detection of timestamp rumor centrality, as a function of $d$, for $\theta = 1$. 
The estimator is run at time $t=d+1$. 
Even at such small timestamps, the probability of detection is empirically close to the upper bound in \eqref{eq:ub}.

%

\vspace{0.07in}
\noindent These results highlight a simple but important point: estimators that exploit graph structure can significantly increase the accuracy of an estimator, up to order-level gains. 


\section{Analysis of Diffusion (Post-2015)}
\label{sec:diffusion}
Having studied the anonymity properties of trickle spreading, we now move to diffusion, a spreading mechanism adopted by the Bitcoin community in 2015 in response to the attacks demonstrated on the trickle spreading mechanism \cite{bitcoindTrickleDiffusion}.
We wish to understand whether diffusion has better anonymity properties than trickle propagation. 
However, the main challenge is that even on simple graphs like a line, computing the exact likelihood of a candidate diffusion source is intractable as the set of infected nodes grows. 
This is because a node's likelihood of being the source depends on the unobserved times at which each node is infected---a state space that grows exponentially in the graph size.
As such, it is challenging to characterize the ML probability of detection in closed form.
Instead, we will compute lower bounds on the ML probability of detection by analyzing two achievable estimation schemes: the first-timestamp estimator 
and a heuristic estimator of our own creation, which we call the \emph{reporting center estimator}.
We will see that these two estimators are complementary, in the sense that they detect the source well for different regimes of degree $d$.
Moreover, they give detection probabilities of the same order as trickle spreading.

\subsection{First-timestamp estimator}
\label{sec:diff_first}
Although the first-timestamp estimator does not use knowledge of the underlying graph, its performance depends heavily on the underlying graph structure. 
The following theorem exactly characterizes its eventual probability of detection on a regular tree.
\begin{theorem} \label{thm:fs_diffusion}
Consider a diffusion process of rate $\lambda=1$ over a $d$-regular tree, $d>2$. Suppose an adversary observes each node's infection time with an independent,  exponential delay of rate $\lambda_2 = \theta$, $\theta \geq 1$. 
Then the following expression describes the probability of detection for the first-timestamp estimator at time $t=\infty$:
\begin{eqnarray}
\prob(\mfs(\boldsymbol \tau, G) = \vs) = \frac{\theta}{d-2}\log\left ( \frac{d+\theta - 2}{\theta}\right  )
\label{eq:fs_diffusion}
\end{eqnarray}
\end{theorem}
(Proof in Section \ref{sec:proof_fs_diffusion})

The proof follows from writing out $\prob(\tau_0 < \tmin)$, conditioned on all the unobserved infection times. 
This expression can be written recursively, which admits a nonlinear differential equation that can be solved exactly. 
The expression highlights a few points:

\vspace{0.07in}
\noindent (1) For a fixed degree $d$, the probability of detection is strictly positive as $t\rightarrow \infty$.
This observation is straightforward in our case (i.e., under the eavesdropper adversary), but under different adversarial models (e.g., snapshot adversaries) it is not trivial to see that the probability of detection
is positive as $t\rightarrow \infty$. Indeed, several papers are dedicated to making exactly that point \cite{SZ11a,SZ12}. 

\vspace{0.07in}
\noindent (2) There is a law of diminishing returns with respect to $\theta$: for a fixed degree $d$, as $\theta$ increases, the rate of growth of \eqref{eq:fs_diffusion} decreases. 
Since $\theta$ represents the number of adversarial connections per honest node, the adversary reaps the largest gains from the first few connections it establishes per node.

\vspace{0.07in}
\noindent (3) When $\theta = 1$, i.e., the adversary has only one connection per node, the probability of detection approaches $\log(d)/d$ asymptotically in $d$. 
Note that this quantity tends to 0 as $d\rightarrow \infty$, and it is order-equal to the probability of detection of the first-timestamp adversary on the trickle protocol when $\theta = 1$ (cf. equation \eqref{eq:trickle1d}).

\vspace{0.07in}
\noindent Theorem \ref{thm:fs_diffusion}  suggests that the Bitcoin community's transition from trickle spreading to diffusion spreading does not provide any order-level anonymity gains (asymptotically in the degree of the graph), at least for the first-timestamp adversary. 
Next, we would like to see whether the same is true for estimators that use the graph structure. 

\subsection{Centrality-Based Estimators}
We compute a different lower bound on the ML probability of detection by analyzing a centrality-based estimator. 
Unlike the first-timestamp estimator, this \emph{reporting centrality estimator} uses the structure of the infected subgraph 
by selecting a candidate source that is close to the center (on the graph) of the observed timestamps. 
However, it does not explicitly use the observed timestamps.
Also unlike the first-timestamp estimator, this centrality-based estimator improves as the degree $d$ of the underlying tree increases.
Indeed, it has a strictly positive probability of detection as $d\rightarrow \infty$, 
implying that the eavesdropper adversary has an ML probability of detection that scales as $\Theta(1)$ in $d$.
We start by presenting the reporting centrality estimator, after which we analyze its probability of detection.

\noindent \textbf{Reporting centrality estimator.} At a high level, the reporting centrality estimator works as follows: for each candidate source $v$, the estimator counts the number of nodes that have reported to the adversary from each of the node $v$'s adjacent subtrees.
It picks a candidate source for which the number of reporting nodes is approximately equal in each subtree.

To make this precise, we introduce some notation. 
First, suppose the infected subtree $G_t$ is rooted at $w$; 
we use  $T^w_v$ to denote the subtree of $G_t$ that contains $v$ and all of $v$'s descendants, with respect to root node $w$.
Consider a random variable $Y_v(t)$, which is 1 if node $v\in V$ has reported to the adversary by time $t$, and 0 otherwise.
We let $Y_{T^w_v}(t) = \sum_{u\in T^w_v}Y_u(t)$ denote the number of nodes in ${T^w_v}$ that have reported to the adversary by time $t$.
We use $Y(t) = \sum_{v \in V_t} Y_v(t)$ to denote the total number of reporting nodes in $G_t$ at time $t$.
Similarly, we use $N_{T^w_v}(t)$ to denote the number of \emph{infected} nodes in ${T^w_v}$ (so $N_{T^w_v}(t) \geq Y_{T^w_v}(t)$), 
and we let $N(t)$ denote the total number of infected nodes at time $t$ ($N(t) \geq Y(t)$).

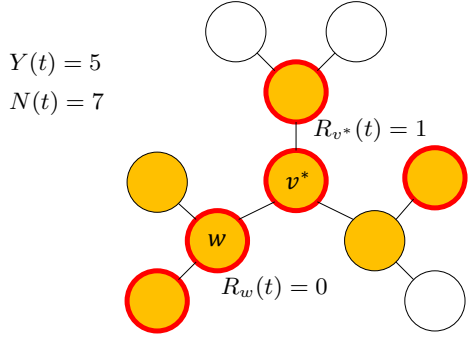
\begin{figure}[t]
    \centering
  \includegraphics[width=.27\textwidth]{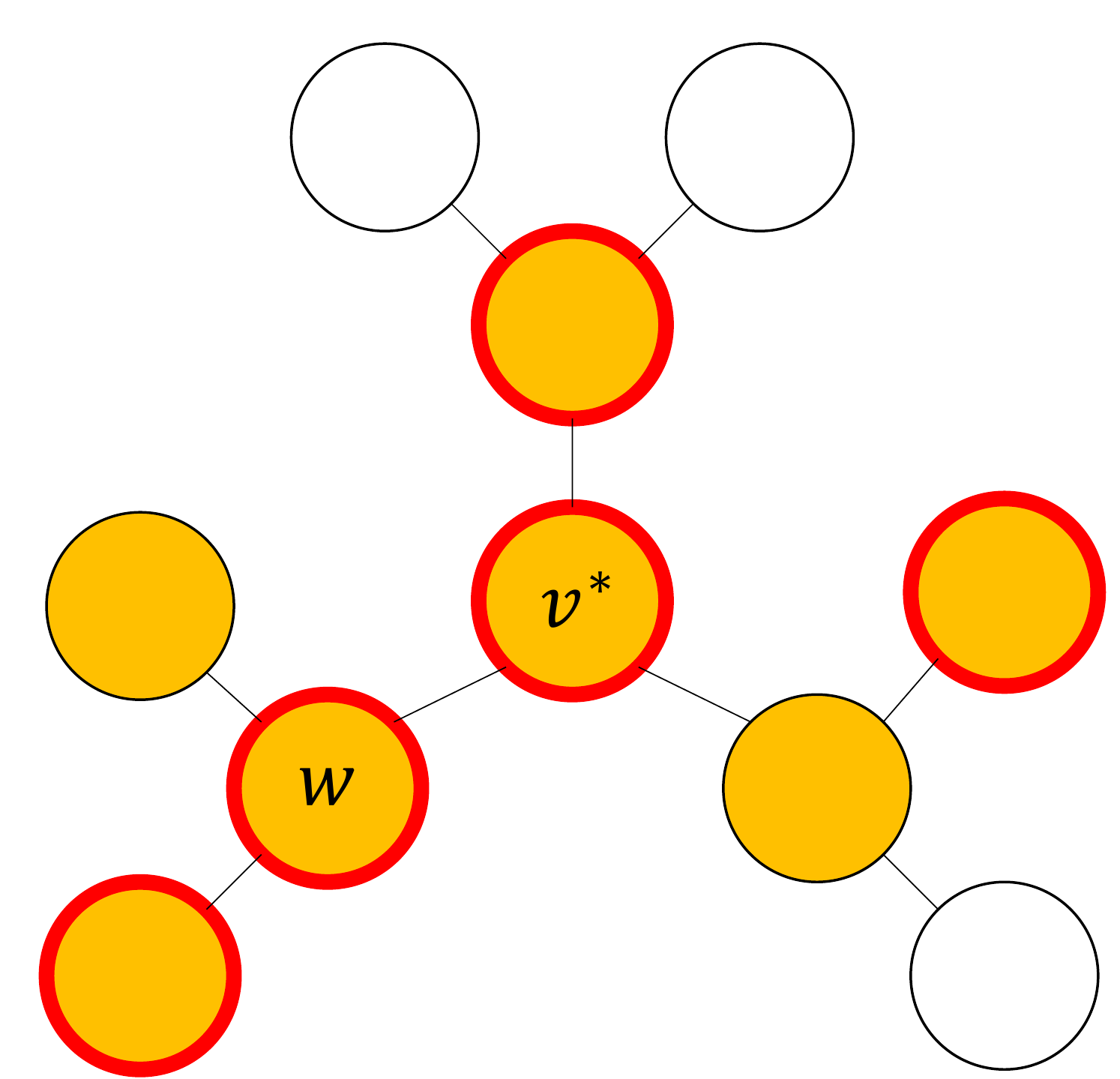}
  \put(-60,75){$R_{\vs}(t)=1$}
  \put(-95,15){$R_{w}(t)=0$}
  \put(-175,100){$Y(t) = 5$}
  \put(-175,85){$N(t) = 7$}
  \caption{Reporting centrality example. Yellow nodes are infected; a red outline means the node has reported. 
  In this example, $\vs$ has $R_{\vs}(t)=1$ because each of its adjacent subtrees has fewer than $Y(t)/2=2.5$ reporting nodes.}
  \label{fig:reporting}
\end{figure}

For each candidate source $v$, we consider its $d$ neighbors, which comprise the set $\mathcal N(v)$. 
We define a node $v$'s \emph{reporting centrality} at time $t$---denoted $R_v(t)$---as follows:
\begin{eqnarray}
R_v(t) = 
\begin{cases} 
      1 & \text{if } \max_{u \in \mathcal N(v)} Y_{T^v_u}(t) < \frac{Y(t)}{2} \\
      0 & \text{otherwise}.
   \end{cases}
\end{eqnarray}
That is, a node's reporting centrality is 1 if and only if each of its adjacent subtrees has strictly fewer than $Y(t)/2$ reporting nodes. 
We say a node is a \emph{reporting center} iff it has a reporting centrality of 1.
The estimator outputs a node $\hat v$ chosen uniformly  from the set of reporting centers.
For example in Figure \ref{fig:reporting}, there is only one reporting center, $\vs$.

Notice that reporting centrality does not use the adversary's observed timestamps---it only counts the number of reporting nodes in each of a node's adjacent subtrees. 
This  estimator is inspired by \emph{rumor centrality}  \cite{SZ11a},  an ML estimator for the source of a diffusion process under a snapshot adversary. 
Recall that a snapshot adversary sees the infected subgraph $G_t$ at time $t$, but it does not learn any timestamp information.

Rumor centrality exhibits a key property on trees: 
a node $v$ is a rumor center of a tree-structured infected subgraph $G_t$ if and only if each of the $d$ adjacent subtrees adjacent has no more than $N(t)/2$ nodes in it \cite{SZ11a}:
\begin{equation}
\max_{u\in \mathcal N(v)}  N_{T^v_u}(t) \leq \frac{N(t)}{2}.
\label{eq:rc_count}
\end{equation}
Moreover, there exists at least one and at most two rumor centers in a tree, and the true source has a strictly positive probability of being a rumor center on regular trees and geometric random trees  \cite{SZ11a}.
In our case, we cannot use rumor centrality directly because the true infected nodes are not all observed at any point in time $t$.
We construct the reporting centrality estimator by applying a condition like \eqref{eq:rc_count} to the \emph{reporting} nodes, rather than the infected nodes.

\noindent \textbf{Analysis.} 
We show that for trees with high degree $d$, reporting centrality has a strictly higher (in an order sense) probability of detection than the first-timestamp estimator; 
its  probability of detection is strictly positive as $d\rightarrow \infty$. 

\begin{theorem} \label{thm:rc_diffusion}
Consider a diffusion process of rate $\lambda=1$ over a $d$-regular tree. Suppose this process is observed by an eavesdropper adversary, which sees each node's timestamp with an independent exponential delay of rate $\lambda_2 = \theta$, $\theta \geq 1$. Then the reporting centrality estimator has a (time-dependent) probability of detection $\prob(\mrc(\boldsymbol \tau, G) = \vs)$ that satisfies
\begin{equation}
\liminf_{t\rightarrow \infty} \prob(\mrc(\boldsymbol \tau, G) = \vs) \geq C_{d} > 0.
\label{eq:const_pd}
\end{equation}
where 
$$
C_{d} = 1-d\left (1-I_{1/2}\left ( \frac{1}{d-2},1+\frac{1}{d-2}\right ) \right )
$$ 
is a constant that depends only on degree $d$, and $I_{1/2}(a,b)$ is the regularized incomplete Beta function, i.e., the probability a Beta random variable with parameters $a$ and $b$ takes a value in $[0,\frac{1}{2})$. 
\end{theorem}
\emph{Proof Sketch: (Full proof in Section \ref{sec:proof_rc_diffusion})}
The proof begins by conditioning on the event that the true source 
is a reporting center.
We demonstrate that this occurs with a probability that is lower-bounded by a constant.
Conditioned on this event, the probability of detection is 1, since there can exist at most one reporting center.
Combining all these observations gives the lower bound in the claim.

The key step in this proof is demonstrating that the source is a reporting center with probability lower-bounded by a constant.
To show this, we relate two P\`olya urn processes: one that represents the diffusion process over the regular tree of honest nodes, and one that describes the full spreading process, which includes both diffusion over the regular tree and random reporting to the adversary.
The first urn can be posed as a classic P\`olya urn  \cite{eggenberger1923statistik}, which has been studied in the context of diffusion \cite{SZ12,khim2015confidence}.
However, the second urn can be described by an unbalanced generalized P\`olya urn (GPU) with negative coefficients---a class of urns that does not typically appear in the study of diffusion (to the best of our knowledge).  
In GPUs, upon drawing a ball of a given color, one can add balls of a different color or even remove balls;
`unbalanced' means that the number of balls added in each timestep is not necessarily equal.
Prior results by Athreya and Ney \cite{athreya2012branching} and later Janson \cite{janson} characterize the limiting distribution of unbalanced GPUs with negative coefficients, 
which allows us to relate the reporting centrality of our process to the rumor centrality of the underlying (unobserved) diffusion process.

\vspace{0.1in}
Notice that the constant $C_d$ in  Theorem \ref{thm:rc_diffusion}  does not depend on $\theta$---this is because the reporting centrality estimator makes no use of timestamp information, so the noisy delays in the observed timestamps do not affect the estimator's asymptotic behavior. 
The delay \emph{does} affect the convergence rate of the probability of detection. 
Characterizing this dependency is theoretically interesting, but beyond the scope of this paper. 

\vspace{0.07in}
\noindent \textbf{Simulation results}.
To evaluate the tightness of the lower bound in Theorem \ref{thm:rc_diffusion}, we simulate reporting centrality on diffusion processes over regular trees.
Figure \ref{fig:rc_vs_fs} illustrates the empirical performance of reporting centrality averaged over 4,000 trials, compared to the theoretical lower bound on the liminf.
The estimator is run at time $t=d+2$.
Our simulations are run up to degree $d=5$ due to  computational constraints, since the infected subgraph grows exponentially in the degree of the tree.
We observe that by degree $d=5$, reporting centrality reaches the theoretical lower bound on the limiting detection probability.

\subsection{First-Timestamp vs. Centrality}
Neither of the lower bounds from the first-timestamp or reporting centrality estimators strictly outperforms the other. 
The first-timestamp estimator performs better on graphs with low degree $d$, whereas reporting centrality performs better in the high-$d$ regime.
By taking the maximum of these two estimators, 
we obtain a lower bound on the ML probability of detection for diffusion processes across the full range of degrees. 
Fully characterizing this probability of detection remains an interesting open problem for future work.

\begin{figure}[t]
    \centering
  \includegraphics[width=.43\textwidth]{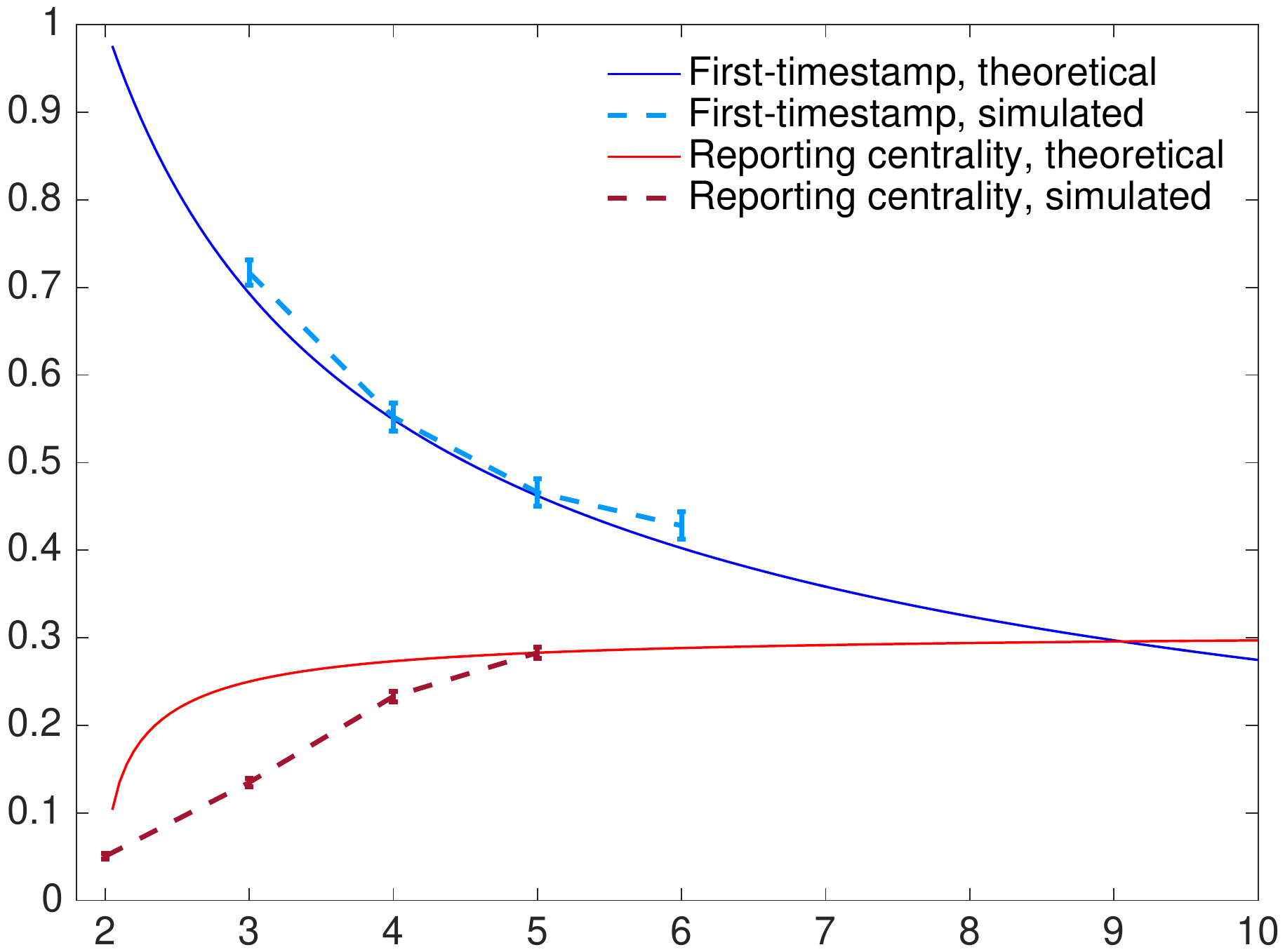}
  \put(-130,-10){Degree, $d$}
  \put(-230, 30){\rotatebox{90}{Probability of Detection}}
  \caption{Comparison of first-timestamp and reporting centrality estimators on diffusion over regular trees,  theoretically and simulated. Here $\theta=\frac{1}{d}$ and $t=d+2$.}
  \label{fig:rc_vs_fs}
\end{figure}

Figure \ref{fig:rc_vs_fs} compares the two estimators both in simulation and theoretically as a function of degree $d$. 
We observe that reporting centrality outstrips first-timestamp estimation for trees of degree 9 and higher;
since our theoretical result is only a lower bound on the performance of reporting centrality, the transition may actually occur at even smaller $d$.
Empirically, the true Bitcoin graph is approximately 8-regular \cite{coinscope},  a regime in which we expect reporting centrality to perform similarly to the first-timestamp estimator.   
Since the practical attacks in \cite{biryukov,koshy2014analysis} use the suboptimal first-timestamp estimator, they may underestimate the ML probability of detection.

\section{Discussion}
\label{sec:experiments}

In this section, we provide a comprehensive comparison of trickle and diffusion, both theoretically and in simulation. 
We also highlight some mathematical subtleties in our analyses, which give rise to a number of open problems of independent interest.

\subsection{Comparison: Trickle vs. Diffusion}
For direct comparison, Table \ref{tab:results_full} contains our theoretical results side-by-side, for the special case of $\theta =1$ and for general $\theta$.
As we have previously stated, the performance of trickle and diffusion are similar, particularly when $\theta =1$.
Although the maximum-likelihood results are difficult to compare at first glance, the point is that they both approach a positive constant as $d,t\rightarrow \infty$; 
for trickle propagation, that constant is $\frac{1}{2}$, whereas for diffusion, it is approximately 0.307 (Corollary 1 from \cite{SZ12}).

\begin{table*}[t]
\centering
\caption{Summary of  probability of detection results on a network of honest servers in  a $d$-regular tree topology. The adversary has $\theta$ connections to each honest server. }
\label{tab:results_full}
\begin{tabular}{|c|c|c|c|}
\hline
\multicolumn{2}{|l|}{}                     & {\bf Trickle} & {\bf Diffusion} \\ \hline
\multirow{2}{*}{\begin{tabular}{@{}c@{}} {\bf First-} \\ {\bf Timestamp} \end{tabular}} & All $\theta$ &   $ \frac{\theta}{d\log 2} \left[\text{Ei}(2^d \log \rho) -\text{Ei}\left ( \log \rho \right ) \right ]$  ~ (Thm \ref{thm:fs_trickle})   &  $\frac{\theta}{d-2}\log\left ( \frac{d+\theta - 2}{\theta}\right  )$ ~ (Thm. \ref{thm:fs_diffusion})      \\ \cline{2-4} 
                           & $ \tiny{\theta = 1} $    &  $\frac{\log(d)}{d \log(2)} + o\left( \frac{\log d}{d}\right)$ ~ (Eq. \ref{eq:trickle1d})        &    
                           		$\frac{\log(d-1)}{(d-2)}$ ~ (Thm. \ref{thm:fs_diffusion})      \\ \hline
\multirow{2}{*}{\begin{tabular}{@{}c@{}} {\bf Maximum-} \\ {\bf Likelihood} \end{tabular}} & All $\theta$ &   $ 1 - \frac{d}{2(\theta + d)}$  ~ (Thm \ref{thm:ml_trickle})   &  $1-d\left (1-I_{1/2}\left ( \frac{1}{d-2},1+\frac{1}{d-2}\right ) \right )$       \\ \cline{2-3} 
                           & $ \tiny{\theta = 1} $    &  $ 1 - \frac{d}{2(d + 1)}$ ~ (Thm. \ref{thm:ml_trickle})        &    
                           		(Thm. \ref{thm:rc_diffusion})        \\ \hline
\end{tabular}
\end{table*}

\begin{figure}[t]
    \centering
  \includegraphics[width=.43\textwidth]{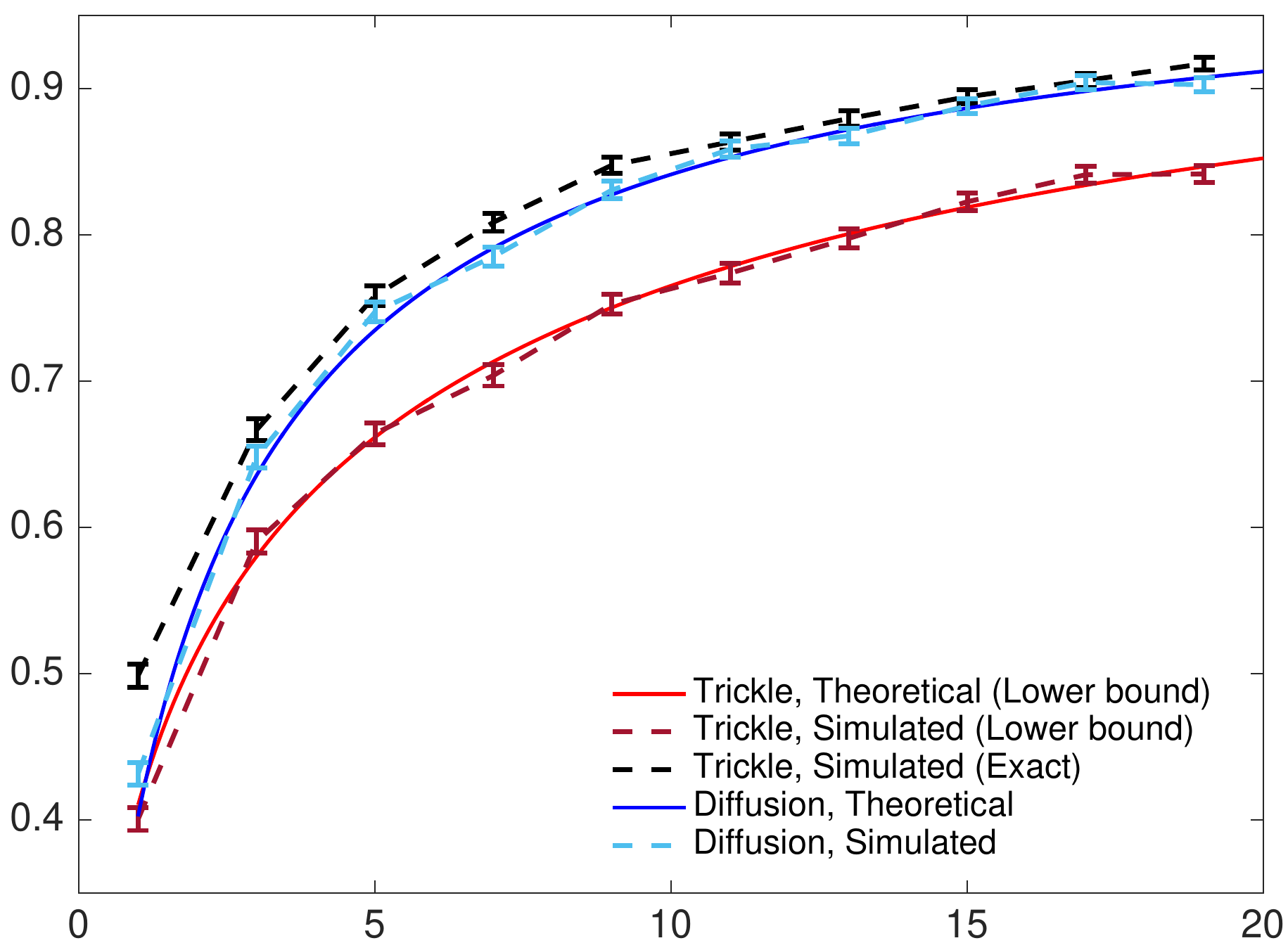}
  \put(-175,-10){Number of corrupted connections, $\theta$}
  \put(-230, 30){\rotatebox{90}{Probability of Detection}}
  \caption{Comparison of trickle and diffusion under the first-timestamp estimator on $4$-regular trees.}
  \label{fig:trickle_vs_diffusion}
\end{figure}

Up to now, we have primarily considered the performance of trickle and diffusion as a function of degree $d$.
However, in practice, the underlying Bitcoin graph is fixed; 
the only variable quantity is the adversary's resources, which are captured by $\theta$, the number of corrupted connections per honest node.
However, it does not make sense to study the asymptotic probability of detection as $\theta \rightarrow 0$, since $\theta\geq 1$; 
on the other hand, as $\theta \rightarrow \infty$, the probability of detection tends to 1, regardless of spreading protcol.
We therefore wish to numerically understand, for a fixed $d$, how diffusion and trickle compare as a function of $\theta$.
Figure \ref{fig:trickle_vs_diffusion} compares analytical expressions and simulation results for the first-timestamp estimator on 4-regular trees. 

We use the first-timestamp estimator for a few reasons: 
first, we lack an efficient ML estimator for diffusion.
Second, the transition from trickle to diffusion was motivated by the practical attacks in \cite{biryukov,koshy2014analysis}, which used a version of the first-timestamp estimator.
Third, even under the first-timestamp estimator, the probability of detection is unacceptably high. 

Note that our \emph{theoretical} results suggest a higher probability of detection for diffusion than for trickle. 
This is an artifact of our lower bound on the trickle probability of detection. 
Recall that our lower bound computes the probability that  $\vs$ reports strictly before any other node (i.e., simultaneous events are discarded). 
Since Figure \ref{fig:trickle_vs_diffusion} is plotted for a small degree $d=4$, simultaneous reporting occurs frequently, so the lower bound is loose.
In simulation, we find that trickle and diffusion actually exhibit nearly identical performance, which agrees with the theoretical probability of detection for diffusion.
Moreover, we verified our trickle lower bound in simulation by discarding realizations where multiple nodes report simultaneously at the first timestamp; these simulation results (burgundy line in Figure \ref{fig:trickle_vs_diffusion}) align closely with our theoretical prediction.
We also find that as $d$ increases, the gap between our trickle lower bound and the simulated probability of detection for trickle decreases. 
Meanwhile, the similarity between diffusion and trickle (in simulation) persists even for high $d$---at least on regular trees.

\begin{figure}[t]
    \centering
  \includegraphics[width=.43\textwidth]{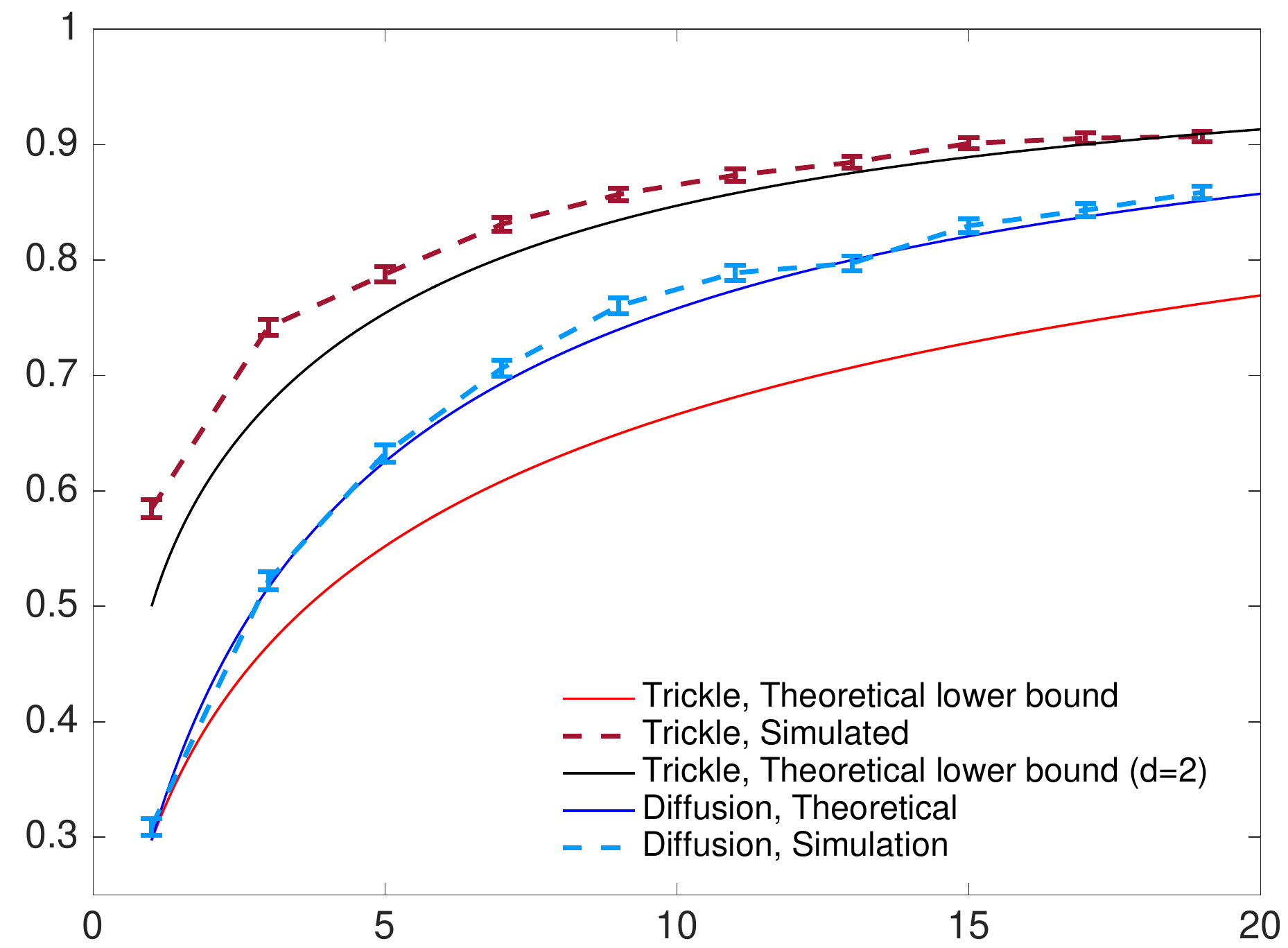}
  \put(-175,-10){Number of corrupted connections, $\theta$}
  \put(-230, 30){\rotatebox{90}{Probability of Detection}}
  \caption{Comparison of trickle and diffusion spreading under the first-timestamp estimator, simulated on a 2015 snapshot of the real Bitcoin network \cite{coinscope}.}
  \label{fig:realgraph_sim}
\end{figure}

\vspace{0.07in}
\noindent \textbf{Real Bitcoin graph.}
The real Bitcoin network is not a regular tree. 
To validate our decision to analyze regular trees, we simulate trickle and diffusion propagation over a snapshot of the real Bitcoin network from 2015 \cite{coinscope}. 
Figure \ref{fig:realgraph_sim} compares these results as a function of $\theta$, for the first-timestamp estimator. 
Since the Bitcoin graph is not tree-structured, we lack ML estimators for both diffusion and trickle; hence, the first-timestamp estimator is a reasonable choice.
Unless specified otherwise, the theoretical curves are calculated for a regular tree with $d=8$, since this is the mean degree of our dataset. 

We first observe that the simulated performance of diffusion is close to our theoretical prediction.  
This occurs because with high probability, the first-timestamp estimator uses only on a local neighborhood to estimate $\vs$. 
Since the Bitcoin graph can be approximated by a sparse, random, regular graph, the graph is locally tree-like with high probability, 
so our theoretical analysis of regular trees applies.
However, our trickle lower bound remains loose.
This is partially due to simultaneous reporting events, but the main contributing factor seems to be the irregularity of the underlying graph.
It appears that trickle responds more acutely to graph irregularities than diffusion, presumably due to its more structured nature, which we exploited heavily in our analysis.
Indeed, we find that the simulated performance of trickle is close to the theoretical predictions for regular trees of degree $d=2$ (black line in Figure \ref{fig:realgraph_sim}); 
$d=2$ happens to be
the mode of the dataset degree distribution.
Understanding this effect more carefully is an interesting question for future work.

Second, we observe that empirically, the probability of detection for diffusion is indeed lower than that of trickle spreading.
This suggests that the Bitcoin developers' intuition was correct---diffusion has  slightly better anonymity properties than trickle. 
Still, the difference is small.
Notice that in Figure \ref{fig:realgraph_sim}, diffusion and trickle become increasingly similar as the number of corrupted connections increases.
In practical attacks \cite{biryukov}, the eavesdropper was able to establish as many as 50 connections to some nodes.
In that regime, Figure \ref{fig:realgraph_sim} suggests that trickle and diffusion would have very similar (high) probabilities of detection, even for the suboptimal first-timestamp estimator. 
Thus, even in a numeric sense, diffusion and trickle are not substantially different under practical adversarial operating conditions. 


In summary, we find that trickle and diffusion have probabilities of detection that are similar, both in an asymptotic-order sense and in a numeric sense, as seen in both simulation and theoretical analysis.
We have evaluated this on the canonical class of $d$-regular trees and, through simulation, on a real Bitcoin graph topology.

\subsection{Open Problems}
Several of our analyses suggest open problems that are interesting in their own right.
For instance, in our analysis of diffusion, the exact ML probability of detection remains unknown.
Moreover, it is unclear if an efficient (polynomial-time in the number of nodes) ML estimator exists.  
Even our analysis of reporting centrality---a suboptimal estimator---gives only a lower bound on the limiting probability of detection, and its convergence rate as a function of $\theta$ and $d$ is not known.
These open problems are all directly related to the goals of this paper. 

\vspace{0.1in}
\noindent \textbf{Relation to other adversarial models.} 
Another subclass of open problems relates to connecting the eavesdropper adversary to more well-known, canonical adversarial models.
That is, the eavesdropper adversary can be thought of as a generalization of certain adversarial models, like the spy-based adversary and the snapshot adversary.

Recall that in the spy-based adversary, each node other than the source is corrupt with probability $p$, and corrupt nodes observe the exact timestamp at which they receive the message.
This setting can be represented by an eavesdropper adversary with random observation delays: either the node reports with an instantaneous delay, or it reports with an infinite delay.

On the other hand, in the snapshot adversary, the adversary learns only which nodes are infected at observation time $t$, not their timestamps.
This is similar to an eavesdropper adversary in which the random reporting delays have high variance. 
As such, the  timestamp has little value beyond indicating that the reporting node is infected.

The similarities between the eavesdropper adversary and these canonical models may be useful to the extent that we can study one to gain intuition about the other.
For example, we can use techniques developed for the eavesdropper adversary to study the spy-based adversary. 
The ML probability of detection for the spy-based adversary on regular trees has evaded exact analysis \cite{KFSV15}, though several heuristics have been found to work well \cite{PTV12,ZY13}.
A simple lower bound on the ML probability of detection comes from analyzing the first-timestamp estimator for the spy-based adversary \cite{KFSV15}. This gives
\begin{equation} \label{eq:spy}
\liminf_{t\rightarrow \infty} \prob(\mfs(\boldsymbol \tau, G)=\vs) \geq p,
\end{equation}
since with probability $p$, the first node to receive the message is a spy.
Although reporting centrality is a suboptimal source estimator for the eavesdropper adversary, it is straightforward to analyze and can be used to obtain lower bounds on the probability of detection for the spy-based adversary. 

Using similar proof techniques to Theorem \ref{thm:rc_diffusion}, we can lower-bound the spy-based adversary's ML probability of detection as time $t\rightarrow \infty$.
%
\begin{corollary}\label{cor:spy}
Consider a diffusion process of rate $\lambda=1$ over a $d$-regular tree. Suppose this process is observed by an spy-based adversary, which sees the exact timestamp of each node independently with probability $p>0$ (otherwise it sees nothing). Then the reporting centrality estimator has a probability of detection $\prob(\mrc(\boldsymbol \tau, G)=\vs)$ that satisfies
\begin{equation}
\liminf_{t\rightarrow \infty} \prob(\mrc(\boldsymbol \tau, G) = \vs) \geq C_{d} > 0,
\label{eq:const_pd_spy}
\end{equation}
where
\[
C_{d} = 1-d\left (1-I_{1/2}\left ( \frac{1}{d-2},1+\frac{1}{d-2}\right ) \right ).
\]
\end{corollary}
(Proof in Section \ref{proof:spy})

The key point to notice about Corollary \ref{cor:spy} is that \eqref{eq:const_pd_spy} remains positive for a fixed $d$ even as $p\rightarrow 0$; this is in contrast with the previously-known equation \eqref{eq:spy}.
Keep in mind  that we are first taking the limit as $t\rightarrow \infty$ for a fixed $p$, then taking $p\rightarrow 0$. 
It is not possible to exchange the order of these limits; doing so drives the probability of detection to zero.
The reason is that for a finite $t$ (and hence a finite number of infected nodes), as $p \rightarrow 0$, the number of spies in the infected subgraph also tends to zero.
Without any observations, it is impossible to detect the source.
However, for any fixed positive $p>0$, as $t\rightarrow \infty$, eventually the number of spies in each subtree concentrates, so reporting centrality has a nonzero probability of detection.
Indeed, our analysis makes critical use of the fact that as $t\rightarrow \infty$, for a fixed positive $p$, the classical P\`olya urn we use to represent the underlying diffusion process concentrates.

Although the lower bound in \eqref{eq:const_pd_spy} does not depend on $p$, the convergence rate does; again, understanding this convergence rate is of theoretical interest.
More broadly, a deeper understanding of the similarities between the eavesdropper adversary and other canonical adversarial models is needed.

\section{Related Work}
\label{sec:related}

There are two primary categories of related work, which originate from the security and network-analysis communities, respectively. 
The first concerns the anonymity properties of bitcoin (and related cryptocurrencies).
The second studies rumor source detection under canonical adversarial and graph models.

\vspace{0.07in}
\noindent \textbf{Bitcoin and Anonymity.}
Bitcoin has long been known to exhibit poor anonymity properties. 
Many papers have explored the anonymity implications of having a  public blockchain.
In particular, pseudonyms can often be linked together (implying common ownership) with simple, heuristic clustering methods. 
For instance, Bitcoin transactions are allowed to have multiple inputs;  in practice, the multiple input pseudonyms are often owned by the same entity. 
Using this and other simple heuristics, researchers have been able to cluster pseudo-nyms in the wild \cite{androulaki2013evaluating,meiklejohn2013fistful,reid2013analysis,ron2013quantitative,ober2013structure}.
Since some nodes are already deanonymized (e.g., public vendors), transaction patterns can be used to learn the identities of other users.
More recently, researchers have demonstrated attacks on the actual P2P network \cite{biryukov,koshy2014analysis}, as discussed in Section \ref{sec:model}.

In response to blockchain-enabled deanonymization threats, several papers have proposed anonymous alternatives to Bitcoin; these alternatives typically rely on privacy-preserving cryptographic protocols.
For example, a new cryptocurrency called Zcash \cite{zcash} uses the Zerocash protocol \cite{sasson2014zerocash}, which cryptographically masks transactions in the blockchain and proves validity using non-interactive zero-knowledge proofs.
Although Zcash obfuscates more information than Bitcoin, the masked transactions are still broadcast over the P2P network using the same protocols as Bitcoin. 
Hence, an eavesdropper adversary could still learn the IP address that originates each transaction.

Another recently-proposed solution is TumbleBit, an untrusted payment hub that can be used for the anonymous transferral of bitcoins \cite{tumblebit}. 
Unlike Zcash, part of the TumbleBit protocol occurs off-blockchain, which prevents tracing attacks by an eavesdropper adversary.
However, the sender and receiver still conduct blockchain operations to escrow and retrieve bitcoins at the beginning and end of each transaction, respectively; these operations use the usual Bitcoin P2P networking protocol, and could be deanonymized (in principle) by a payment hub that also acts as an eavesdropper adversary.
In summary, several anonymity-preserving altcoins are vulnerable to the kinds of attacks outlined in this paper, despite their strong cryptographic guarantees at the application layer. 

\vspace{0.07in}
\noindent \textbf{Rumor Source Detection.} The last five years have seen significant work on detecting the source of a diffusion process  over a graph.
This topic became popular following the 2010 results of Shah and Zaman, who showed that under a snapshot adversary on a regular tree, one can reliably infer the source of a diffusion process \cite{SZ10}.
That is, the probability of detection is lower-bounded by a constant as $t\rightarrow \infty$.
The analysis for these results uses the same P\`olya urn construction that we adopt in the proofs of Theorem \ref{thm:rc_diffusion} and Corollary \ref{cor:spy} \cite{SZ11a,SZ11b}.
Shah and Zaman later extended these results to random, irregular trees \cite{SZ12}, and other authors studied heuristic source detection methods on general graphs \cite{FC12,PVF12,LMOZ13} and related theoretical limits \cite{WDZT14,milling2012network,khim2015confidence}.

Follow-up research has considered several variants on the snapshot adversary. 
For example, Pinto et al. considered a spy-based adversary that observes a diffusion process where the delays are truncated Gaussian random variables \cite{PTV12}, 
and Zhu and Ying consider a spy-based adversary with standard exponential delays \cite{ZY13}. 
These papers do not characterize the ML probability of detection, but they do propose efficient heuristics that perform well in practice. 
Our work instead analyzes the eavesdropper adversary, a new adversarial model that emerged from practical attacks on the Bitcoin network \cite{biryukov,koshy2014analysis}.
As we saw in various theoretical results, the eavesdropper adversary requires completely new analytical tools compared to the spy-based and snapshot adversaries.

Similarly, there has been extensive work on alternative spreading models. 
In particular, researchers have studied various forms of diffusion, in which nodes can ``recover from the infection"---i.e., they can delete the transaction before it reaches the entire network. 
The classic diffusion process is often called susceptible-infected (SI): nodes start as susceptible, and once they become infected, they remain so for the rest of time.
 Alternative models include susceptible-infected-susceptible (SIS) diffusion, in which nodes recover from the infection with a random delay and can then be re-infected,
and susceptible-infected-recovered (SIR) diffusion, in which nodes recover with an exponential delay, upon which they are immune to future infection.
These models have been studied  in the literature, both theoretically and empirically \cite{ZY14,CZY16,ZY16,BT95}. 
In our work, we consider only classical SI diffusion processes, but we also consider the completely new trickle spreading protocol.
As we have seen in Theorems \ref{thm:fs_trickle}  and \ref{thm:ml_trickle}, the theoretical machinery for analyzing trickle spreading differs significantly from that used to analyze diffusion.
The differences arise from the combinatorial nature of trickle spreading.

\section{Conclusion}
\label{sec:discussion}

In this paper, we analyze the anonymity properties of the Bitcoin P2P network, with particular attention to the shift from trickle to diffusion propagation in 2015. 
We find that trickle and diffusion have similar (poor) anonymity properties. 
On regular trees, they exhibit the same scaling properties asymptotically in $d$, and numerically similar probabilities of detection for a fixed $d$.
In simulation over a real Bitcoin graph topology, diffusion and trickle also exhibit numerically similar performance, which agrees with our theoretical predictions. 
This leads us to conclude that the current flooding protocols used in the Bitcoin network do not sufficiently protect user anonymity. 

An interesting question is how to modify the networking stack in order to provide robustness to source deanonymization attacks.
Although the design of such solutions is beyond the scope of this paper, our analysis gives some intuition for how to prevent deanonymization attacks. 
A key reason that deanonymization is currently possible is because of the symmetry of current spreading protocols. 
That is,  diffusion and trickle  both propagate content over the underlying graph in all directions at roughly the same rate.
This symmetry enables powerful centrality-based attacks. 
Thus, a natural solution is to break the symmetry of diffusion and trickle.
Understanding how to break symmetry without hurting performance is of both  theoretical and practical  interest.

\section*{Acknowledgments}
The authors would like to extend heartfelt thanks to Andrew Miller for pointing out the Bitcoin community's transition from trickle to diffusion, Varun Jog for valuable insights about the limiting behavior of P\`olya urns, and Sewoong Oh for discussions on modeling the evolution of diffusion processes with an eavesdropper adversary.

\bibliographystyle{plain}
\bibliography{privacy}

\clearpage
\appendix

\begin{figure*}[ht]
\caption{{\sc Timestamp Rumor Centrality}. Returns the timestamp rumor centrality of candidate source $v$, given timestamps $\btau$ on tree $G$. 
$\phi(w)$ denotes the children of node $w$ on a tree $T$ rooted at $v$. $\partial T$ denotes the leaves of tree $T$.}
\label{algo:timestamp}
\begin{multicols}{2}
\begin{algorithmic}[1]
\renewcommand{\algorithmicrequire}{\textbf{Input:}}
\renewcommand{\algorithmicensure}{\textbf{Output:}}
\Require Timestamps $\btau$, graph $G$, candidate source $v$, estimation time $t \geq d+1$
\Ensure Number feasible orderings from source $v$, $f_v$
\State $[d+1] \triangleq \{1,\ldots, d+1\}$
\State $m_{vv}\gets \{0\}$
\State \Comment Two global variables, $T$ and $\mathcal C$:
\State $T \gets$ balanced tree of radius $t$, center $v$, over graph $G$
\State $\mathcal C \gets$ dictionary of number of feasible partial orderings for each feasible ($w$, $\hat X_w$) pair
\State $c = ${\sc PassToLeaves}($v,v, \{0\}, T$)
\Return $c$ 
\Comment Timestamp rumor centrality for $v$
\State
\Function{PassToLeaves}{$z,w,m, T$}
    \Comment Each node learns its possible receipt times, recursively
    \State $m(w)\gets m \setminus \{t\in m: t\geq \tau_w\}$
    \Comment Remove late times \;
    \State Node $w$ saves $m(w)$ outside function scope
    \If {$w \in \partial T$}
    	\For {$t' \in m$}
	    	\State $\mathcal C[w][t'] \gets 1$
		\Comment Number of partial orderings with $w$ infected at time $t'$
	\EndFor
    \Else
    	\State $m'\gets \cup_{i \in m(w)} \{i + j: j\in [d+1]\} $
    	\State $m' \gets m' \setminus \{\tau_w\}$ 
	 \For {$y \in \phi(w)$}
    		\State {\sc PassToLeaves}$(w,y,m')$
    	\EndFor
    \EndIf
    \State {\sc AggregateMessages}($w,\phi(w)$)
    \If{$z = w = \vs$}
	    \Return $\mathcal C[\vs][0]$
    \EndIf
\EndFunction
\State
\Function{AggregateMessages}{$z,\phi(z),T$}
\Comment Counts the number of valid orderings by passing messages that represent the set of feasible orderings
    \State $\mathcal N \gets [z, \phi(z)] $
    \Comment Ordered list of parent and children
    \State $M\gets \prod_{u \in \phi(z)}m(u)$ 
    \State
    \Comment $\prod$ denotes Cartesian set product, where $A\times B\triangleq \{(a,b):a\in A, b\in B\}$
    \State $M \gets m(w)\times M $
    \Comment Prepends the current node's feasible receipt times to the Cartesian product
    \State $M \gets \{m\in M: [(m_1,\ldots,m_{d}) \text{ distinct}] \land [|m_i-m_j|\leq d+1$ for all $i,j$]  $\land [m_1 < m_i$ for all $i>1] \}$
    \State
    \Comment Removes ordered tuples where neighbors receive the message at the same time, are too far apart to be feasible, or parent gets message after children. $m_i$ denotes $i$th element of ordered tuple $m$
    \For{m $\in M$}
    	\State $q\gets \prod_{i=2}^d \mathcal C[\mathcal N_i][m_i]$
	\Comment Compute the number of permutations by multiplying counts from each child node
	\State $\mathcal C[z][m_1] \gets C[z][m_1] + q$
    \EndFor
    \Return
\EndFunction
\end{algorithmic}
\end{multicols}
\end{figure*}

\section{Proofs}
\label{app:proofs}

\section{Algorithms}
\label{app:algorithms}

Timestamp rumor centrality is described in Protocol \ref{algo:timestamp}.
Here we have assumed that the adversary has only one link per honest server, so $\theta = 1$. 
This is easily extended to general $\theta$.


\subsection{Proof of Theorem \ref{thm:fs_trickle}}
\label{proof:fs_trickle}
We can write this lower bound explicitly. 
Recall that the system is discrete-time, and each node has $d$ honest neighbors and $\theta$ connections to the adversary. 
$\tau_i$ denotes the \emph{first} time node $i$ reports to the adversary.
\begin{eqnarray}
\prob(\tau_0 < \tau_m) = \sum_{i=1}^{d+\theta} \prob (\tau_0=i) \prob(\tau_m > i | \tau_0=i)
\end{eqnarray}
Since the source has only $d$ honest connections, for any $i > (d+1)$, it holds that $\prob(\tau_0=i)=0$, and if $i=d+1$, then $\tau_m \leq i$.
So we can simplify this summation to
\begin{eqnarray}
\prob(\tau_0 < \tau_m) = \sum_{i=1}^{d} \prob (\tau_0=i) \prob(\tau_m > i | \tau_0=i).
\end{eqnarray}
This means we only need to consider the first $d$ time steps of the message spread in order to lower-bound the first-timestamp adversary's probability of detection.
Let 
$
a_j \triangleq \prob (\tau_m = j| \tau_0 \geq j)=\prob (\tau_m = j| \tau_0 = i \text{ for any } i\geq j).
$
Then $\prob(\tau_m > i | \tau_0 = i) = 1 - \sum_{j=1}^i a_j,$
so
\begin{align}
\prob(\tau_0 < \tau_m) &=& \sum_{i=1}^d \prob(\tau_0=i) - \sum_{i=1}^d \prob(\tau_0=i) \sum_{j=1}^i a_j \nonumber \\
&=& 1-\prob(\tau_0 = d+1) - \sum_{i=1}^d \prob(\tau_0=i) \sum_{j=1}^i a_j. \label{eq:pd_ft}
\end{align}
Letting
$
b_k \triangleq \prob (\tau_m > k ~|~ \tau_0 \geq k, \tau_m > (k-1))
$
gives 
\begin{eqnarray}
a_j = (1-b_j) \prod_{k=1}^{j-1}b_k.
\label{eq:aj}
\end{eqnarray}
We can write $b_k$ explicitly by noting that as long as no infected nodes have reported the message to the adversary, one can deterministically compute the number of infected nodes at each time step with a given number of infected (resp. uninfected) neighbors.
This is because the underlying graph is a regular tree (see Lemma \ref{lem:ordering} for proof).
Let $\mathscr N(k,t)$ denote the set of nodes with $k$ infected, honest neighbors at time $t$. Then for a fixed time $t$, we can compute the probability that \emph{every} infected node chooses to infect an honest node in the next time step, by indexing over the value of $k$:
\begin{eqnarray*}
b_j &=& \prod_{k=1}^{j-1} \left ( \frac{ |\{v\in V: u_{v,h}(j)=k\}|}{|\{v\in V: u_v(j)=k\}|}\right)^{|\mathscr N(k,j)|} \\
&=& \prod_{k=1}^{j-1} \left (\frac{d-k}{d-k + \theta } \right ) ^{2^{j-1-k}}
\end{eqnarray*}
where $j > 1$, and $b_1 = 1$.
Here, the quantity $u_{v}(j)$ (resp. $u_{v,h}(j)$) denotes the uninfected degree (resp. uninfected honest degree) of node $v$ at time $j$---that is, the number of total (resp. honest), uninfected neighbors of a node.
So the ratio in the definition of $b_j$ is comparing the number of nodes with honest uninfected degree $k$ to the number of nodes with uninfected degree $k$.

Substituting this quantity into Equation \eqref{eq:aj}, we get
\begin{eqnarray}
a_j &=& (1-b_j) \prod_{k=1}^{j-2} \prod_{m=1}^k  \left ( \frac{d-m}{d-m+\theta } \right )^{2^{k-m}} \nonumber \\
      &=& (1-b_j) \prod_{m=1}^{j-2} \left ( \frac{d-m}{d-m+\theta } \right )^{\sum_{\ell=0}^{j-m-2}2^{\ell}} \nonumber \\
      &=& (1-b_j) \underbrace{ \prod_{m=1}^{j-2} \left ( \frac{d-m}{d-m+\theta } \right )^{2^{j-m-1}-1} }_{M_j} \label{eq:aj_expanded}
\end{eqnarray}
Rearranging, we get
\begin{eqnarray*}
M_j &=& \frac{ \prod_{m=1}^{j-1}  \left ( \frac{d-m}{d-m+\theta } \right )^{2^{j-1-m}}  \prod_{k=1}^{j-2}\left (\frac{d-k+\theta }{d-k} \right )} {\frac{d-j+1}{d+\theta-j+1}} \\
       &=& b_j \left( \frac{d+\theta-j+1}{d-j+1} \right) \underbrace{\prod_{k=1}^{j-2}\left (\frac{d-k+\theta }{d-k} \right )}_{W_j}
\end{eqnarray*}
Writing out the terms of $W_j$ explicitly, we get that
\begin{eqnarray*}
W_j &=& \frac{(d-1+\theta )(d-2+\theta )\ldots (d-1)\ldots (d-j+2+\theta )}{(d-1)\ldots (d-j+2+\theta )\ldots (d-j+2)}\\
       &=& \prod_{m=0}^{\theta -1}\frac{d+m}{d+m-j+2}.
\end{eqnarray*}
Substituting all of this into Equation \eqref{eq:aj_expanded}, we get that
\begin{eqnarray*}
a_j = (1-b_j)b_j \left ( \frac{d-j+1+\theta }{d-j+1}\right )\prod_{m=0}^{\theta -1}\frac{d+m}{d+m-j+2}.
\end{eqnarray*} 
This expression for $a_j$ can be used to rewrite  \eqref{eq:pd_ft}: 
\begin{eqnarray}
\prob (\tau_0 < \tau_m) &=& 1-\underbrace{\prob(\tau_0 = d+1)}_{P_A} - \sum_{i=1}^d \prob(\tau_0=i) \sum_{j=1}^i a_j. \nonumber \\
           			    &=& 1- P_A - \sum_{i=1}^d \prob(\tau_0=i) \sum_{j=1}^i (1-b_j)b_j  \times \nonumber \\ 
			    && \frac{d-j+1+\theta }{d-j+1} \prod_{m=0}^{\theta -1}\frac{d+m}{d+m-j+2}.
\end{eqnarray}
Setting
$$
q_j \triangleq \frac{d-j+1+\theta }{d-j+1} \prod_{m=0}^{\theta -1}\frac{d+m}{d+m-j+2},
$$
we get that 
\begin{align}
\prob (\tau_0 < \tau_m) &=& 1- P_A - \sum_{i=1}^d \prob(\tau_0=i) \sum_{j=1}^i  (1-b_j)b_j q_j \nonumber \\
				    &=& 1- P_A - \sum_{i=1}^d \prob(\tau_0=i) \left (\sum_{j=2}^i  b_j q_j - \sum_{j=2}^i  b_j^2 q_j \right ) \label{eq:reduce_summation}
\end{align}
where the change of summation bounds in \eqref{eq:reduce_summation} occurs because $b_1=1$.
We define
$
\gamma_k \triangleq \left (\frac{d-k}{d-k+\theta } \right )^{2^{-k}},
$
which means that 
\begin{eqnarray}
b_j = \prod_{k=1}^{j-1}\gamma_k^{2^{j-1}}.
\label{eq:bj}
\end{eqnarray}
Using this notation, we write out the two final summations in \eqref{eq:reduce_summation} explicitly:
\begin{align}
\sum_{j=2}^i  b_j q_j = q_2 \gamma_1^2 + q_3 \gamma_1^4 \gamma_2^4 + q_4 \gamma_1^8 \gamma_2^8 \gamma_3^8 + \ldots + q_i \gamma_1^{2^{i-1}} \cdots \gamma_{i-1}^{2^{i-1}} \label{eq:sum_linear}\\ 
\sum_{j=2}^i  b_j^2 q_j = q_2 \gamma_1^4 + q_3 \gamma_1^8 \gamma_2^8 + q_4 \gamma_1^{16} \gamma_2^{16} \gamma_3^{16} + \ldots + q_i \gamma_1^{2^{i}} \cdots \gamma_{i-1}^{2^{i}} \label{eq:sum_square}
\end{align}
Subtracting  \eqref{eq:sum_linear}-\eqref{eq:sum_square} and collecting terms gives
\begin{align}
\sum_{j=2}^i  q_j(b_j-b_j^2) = q_2 \gamma_1^2 +  \gamma_1^4 (q_3 \gamma_2^4 - q_2) +  \gamma_1^8 \gamma_2^8 (q_4 \gamma_3^8 - q_3) + \nonumber \\
				           \ldots +  \gamma_1^{2^{i-1}} \cdots \gamma_{i-2}^{2^{i-1}}(q_i \gamma_{i-1}^{2^{i-1}}-q_{i-1}) - q_i \gamma_1^{2^i}\cdots \gamma_{i-1}^{2^i} \nonumber \\
				           = q_2 \gamma_1^2 - q_i \gamma_1^{2^i}\cdots \gamma_{i-1}^{2^i} + \sum_{\ell=3}^i \gamma_1^{2^{\ell-1}}\cdots \gamma_{\ell-2}^{2^{\ell-1}}(q_\ell \gamma_{\ell-1}^{2^{\ell-1}}-q_{\ell-1}).\label{eq:sum_diff}
\end{align}
First, we show that the summation (last term) in \eqref{eq:sum_diff} is equal to 0 by writing out $(q_\ell \gamma_{\ell-1}^{2^{\ell-1}}-q_{\ell-1})$:
\begin{eqnarray*}
q_\ell \gamma_{\ell-1}^{2^{\ell-1}}-q_{\ell-1} = q_{\ell-1}\left (\frac{d-\ell+1+\theta }{d-\ell+1} \gamma_{\ell-1}^{2^{\ell-1}} - 1\right ) \\
= q_{\ell-1}\left (\frac{d-\ell+1+\theta }{d-\ell+1} \cdot \frac{d-\ell+1}{d-\ell+1+\theta }  - 1\right ) = 0.
\end{eqnarray*}
Next, we show that the first term in \eqref{eq:sum_diff} equals 1 by writing $q_2 \gamma_1^2$ explicitly:
$$
q_2 \gamma_1^2 = \frac{d-2+1+\theta }{d-2+1} \left ( \prod_{m=0}^{\theta  -1}\frac{d+m}{d+m-2+2} \right ) \frac{d-1}{d-1+\theta } = 1,
$$
which implies that 
\begin{eqnarray}
\sum_{j=1}^i  q_j(b_j-b_j^2) = 1 - q_i \gamma_1^{2^i}\cdots \gamma_{i-1}^{2^i} = 1 - b_i^2.
\label{eq:difference}
\end{eqnarray}
Now, we can substitute \eqref{eq:difference} into \eqref{eq:reduce_summation}, getting
\begin{eqnarray}
 \prob (\tau_0 < \tau_m) &=& 1- P_A - \sum_{i=2}^d \prob(\tau_0=i) \big (1 - q_i \gamma_1^{2^i}\cdots \gamma_{i-1}^{2^i} \big ) \nonumber \\
 				     &=& \frac{\theta}{\theta+d} + \sum_{i=2}^d \prob(\tau_0=i) q_i \gamma_1^{2^i}\cdots \gamma_{i-1}^{2^i}, \label{eq:final_simp}
\end{eqnarray}
where \eqref{eq:final_simp} is because 
\begin{align*}
\sum_{i=2}^d \prob(\tau_0=i) &=& 1 - \prob(\tau_0 = d+1) - \prob(\tau_0 = 1) \\
					    &=& 1-P_A - \frac{\theta}{\theta+d}.
\end{align*}
Note that 
$
P(\tau_0 = i) = \frac{{{N-i}\choose{\theta -1}}}{{{N} \choose {\theta}}},
$
where $N \triangleq d+\theta$. We can also rewrite $q_i$ as 
\begin{align*}
q_i = \frac{d-i+1+\theta }{d-i+1}\prod_{m=0}^{\theta -1}\frac{d+m}{d+m-i+2} \\
     = \frac{d-i+1+\theta }{d-i+1} \cdot \frac{(d+\theta  - 1)!}{(d-1)!} \cdot \frac{(d-i+1)!}{(d+\theta +1-i)!}\cdot \frac{\theta !}{\theta!}\\
     = \frac{{{N-1}\choose {\theta }}}{{{N-i }\choose {\theta }}}.
\end{align*}
Thus, the product of these terms is
\begin{eqnarray*}
P(\tau_0 = i)q_i &=& \frac{\theta }{N-i-\theta +1} \cdot \frac{N-\theta }{N} \\ 
			 &=& \frac{\theta}{d-i+1} \cdot \frac{d}{\theta +d},
\end{eqnarray*}
and the desired probability in \eqref{eq:final_simp} simplifies to 
\begin{eqnarray}
 \prob (\tau_0 < \tau_m) &=& \frac{\theta}{\theta+d} + \sum_{i=2}^d \frac{d}{d-i+1} \cdot \frac{\theta}{\theta +d}b_i^2 \nonumber \\
 				     &=& \frac{\theta}{\theta+d}\left [1 + d \sum_{i=2}^d \frac{b_i^2}{d-i+1} \right ] \nonumber \\
				     &\geq& \frac{\theta}{\theta+d}\left [1 +  \sum_{i=2}^d b_i^2 \right ] \label{eq:lastline} \\
 				     &\geq & \frac{\theta}{\theta+d}\left [1 +  \sum_{i=1}^{d-1} \gamma_1^{\sum_{k=1}^i 2^k}\right ] \label{eq:approxd} \\
				     &=& \frac{\theta}{\theta+d}\left [1 +  \sum_{i=1}^{d-1} \gamma_1^{2^{i+1} -2}\right ] \nonumber \\
				     &\geq& \frac{\theta}{d} \sum_{i=0}^{d-1} \gamma_1^{2^{i+1}} \label{eq:larged} 
 \end{eqnarray} 
 where \eqref{eq:lastline} comes from replacing $d-i+1$ with $d$,
 \eqref{eq:approxd} comes from replacing  $\gamma_k$ with $\gamma_1$ since $\gamma_k \geq \gamma_1$, 
and \eqref{eq:larged} holds because $\gamma_1^2 \leq \frac{d}{d+\theta}$. 
We lower-bound the doubly exponential sum in \eqref{eq:larged} by integrating. Letting $\rho=\gamma_1^2$, we have
\begin{eqnarray*}
\frac \theta d  \sum_{i=0}^{d-1} (\gamma_1^2)^{2^i}  &\geq & \frac \theta d \int_0^{d}  \rho^{2^x}dx \nonumber \\
									  &=& \frac{\theta}{d \log 2} \left (\text{Ei}(2^d \log \rho^2) - \text{Ei}( \log \rho^2) \right ) 
 \end{eqnarray*}
 where Ei$(\cdot)$ denotes the exponential integral.  This gives the desired result.

\subsection{Proof of Proposition \ref{lem:ordering}}
\label{proof:ordering}
The proposition can be seen through a simple counting argument.
Let $A_t$ denote the set of \emph{active nodes} at time $t$, or the set of all infected, honest nodes with at least one uninfected neighbor (honest or adversarial).
We also define the \emph{uninfected degree} of a node $u_v(t)$ as the number of uninfected neighbors of node $v$ at time $t$.
We define $A_t(i)=|\{v\in V : u_v(t) = i\}|$, $i>0$, as the number of active nodes at time $t$ with uninfected degree $i$.

We claim that for a given regular tree $G$ and set of observed timestamps $\boldsymbol \tau$, and for any set of feasible orderings, the number of nodes with uninfected degree $i$ (i.e., $A_t(i)$) for a given $i>0$ does not depend on  the underlying ordering. 
This holds because the graph is regular; we can show it formally by induction. 
At time $t=1$, there are two options: either the source reports directly to the adversary, or it spreads the message to an honest neighbor.
It if reports to the adversary, then in every feasible ordering, the source must report to the adversary at $t=1$.
Therefore at the end of time step $t=1$, we have 
$$
A_t(i) = \begin{cases}
1 \qquad \text{if }i = d+\theta - 1\\
0 \qquad \text{otherwise.}
\end{cases}
$$
If the source instead passes the message to an honest neighbor (it doesn't matter which one, and the identity of the neighbor could vary across feasible orderings), then
$$
A_t(i) = \begin{cases}
2 \qquad \text{if }i = d+\theta - 1\\
0 \qquad \text{otherwise,}
\end{cases}
$$
because now we have two active nodes, each of which has $d + \theta  -1$ uninfected neighbors.

Now take $t>1$, and assume that at time $t-1$, $A_t(i)$ was the same across all feasible orderings, for each  $i>0$.
We want to show that the same is true at time $t$.
Every time an active node $v$ infects a neighbor node $w$, $v$'s own uninfected degree decreases by one. 
If $w$ is an honest node, it joins the set of active nodes with uninfected degree $u_w(t) = d-1 + \theta $.
If $w$ belongs to the eavesdropper, then it does not join the active nodes. 
Suppose $\boldsymbol \tau$ indicates that in time $t$, exactly $m$ nodes will report to the adversary.
Since there were $|A_{t-1}|$ active nodes at time $t$, we know that $|A_{t-1}|-m$ new active nodes will be infected, each with degree $d-1+\theta $, 
so $A_t(d-1+\theta ) = |A_{t-1}| -m$. 
Moreover, for all $0<i<(d-1+\theta )$, we have $A_t(i) = A_{t-1}(i+1)$, since each previously-active node decrements its uninfected degree by one.
None of this depends on the ordering, which proves that $A_t(i)$ takes the same value for any feasible ordering. 

We write out the likelihood $L(o)$ of a feasible ordering $o$:
\[
L(o) = \prod_{j = 1}^{t-1} \prod_{v \in A_j} \frac{1}{u_v(j)},
\]
since each active node infects exactly one uninfected neighbor uniformly at random in each time step.
By the previous argument, this likelihood can equivalently be grouped by nodes with the same uninfected degree $u$:
\[
L(o) = \prod_{j = 1}^t \prod_{u \in \{1,\ldots, d-1 + \theta \}} \left ( \frac{1}{u} \right )^{A_j(u)}.
\]
Nothing in this expression depends on the ordering $o$ (since $A_j(u)$ is independent of $o$), so the likelihood must be equal for all feasible orderings.

\subsection{Proof of Theorem \ref{thm:ml_trickle}}
\label{proof:ml_trickle}
We begin by showing that \eqref{eq:ub} is an upper bound on any estimator's probability of detection, and then show that ball centrality achieves the lower bound in \eqref{eq:lb}. 

\noindent \emph{(1):} Notice that with probability $\theta /(\theta+d)$, the true source reports the message to the adversary in the first time slot. 
If this happens, then the conditional probability of detection is 1; since the adversary is assumed to know the starting time of the trickle process, it can deduce that the true source is the only possible source. 
However, if the source does \emph{not} report to the adversary in the first time slot, then the source must instead pass the message to one of its honest neighbors.
In that case, at $t=1$, exactly two nodes are infected, both of which are honest. 
Thereafter, the spread from each of these nodes is completely symmetric: both the true source and the first neighbor
are connected to identical graph structures (i.e., two infinite trees rooted at each of the infected nodes), and the message spreading dynamics from each of these two nodes is identically-distributed. 
As such, no estimator can distinguish between the true source and the first neighbor, meaning the probability of detection 
is upper bounded by $1/2$ in this case.
Therefore, a simple upper bound for the ML probability of detection is
\begin{eqnarray*}
\prob(\mml(\boldsymbol \tau, G) = \vs)  &\leq& \frac{\theta}{\theta + d} \big (1 \big )+ \frac{d}{\theta + d} \left (\frac{1}{2} \right ) \nonumber \\
				    &=& 1 - \frac{d}{2(\theta + d)}.
\end{eqnarray*}

\noindent \emph{(2): } Next, we demonstrate that ball centrality has a probability of detection that is lower bounded by \eqref{eq:lb}.
Once again, if the source immediately passes the message to the adversary, then the adversary identifies a ball of size 1, so the source gets caught with probability 1.
This happens with probability $\frac{\theta}{\theta + d}$. 
Thus we need to compute the probability of detection conditioned on the source passing the message to an honest neighbor at $t=1$.
Suppose source $\vs$ passes the message to its neighbor $w$ at $t=1$.
Without loss of generality, let us think of $w$ and $\vs$ as the left and right respective endpoints of a path.
In each subsequent time step, each of the endpoints of this path will forward the message either to another honest node or to the adversary. 
If the left (resp. right) endpoint $q$ forwards the message to another honest node $z$, then $z$ becomes the left (resp. right) endpoint in the following time step;
if $q$ forwards the message to the adversary, then the path terminates at $q$. 
This path continues to grow until both ends have terminated.

We first show that the probability of the path terminating is lower bounded by $1-\frac{2}{(1+\theta/d)^{t-1}}$, then we show that conditioned on path termination, the probability of detection is lower bounded by $1/2$.
To analyze the probability of termination at time $t$, note that each end of the path terminates after a geometrically-distributed number of hops.
This holds because each endpoint independently terminates with probability $\frac{\theta}{\theta +d-1 }$ in each time step.
Therefore, the probability of both endpoints terminating by time $t$ can be expressed as
\begin{eqnarray}
\prob(B \leq t)^2 &=& \left ( 1 - \left ( 1 - \frac{\theta}{\theta + d-1}\right )^{t-1}\right )^2, \nonumber \\
			&\geq& \left ( 1 - \left (\frac{d}{\theta + d}\right ) ^{t-1} \right )^2 \nonumber \\
			&\geq& 1 - 2\left (\frac{d}{\theta+d} \right )^{t-1} \label{eq:termination}
\end{eqnarray}
where $B$ is a geometric random variable of rate $\frac{\theta}{\theta +d-1 }$.

\begin{figure}[h]
    \centering
  \includegraphics[width=.4\textwidth]{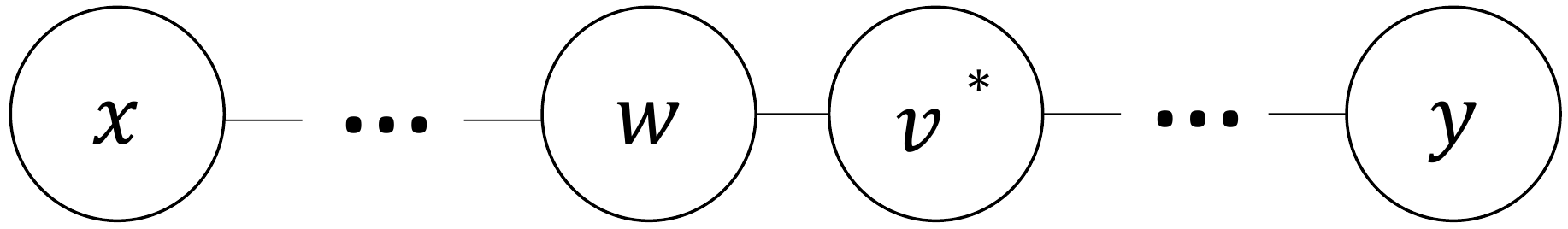}
  \caption{Arrangement of nodes from the proof of Thm. \ref{thm:ml_trickle}.}
  \label{fig:line}
\end{figure}

Now, we show that conditioned on termination, the probability of detection is lower-bounded by $\frac{1}{2}$.
We call $x$ and $y$ left and right terminating endpoints, respectively (Fig. \ref{fig:line}).

Now, we show that the balls centered at nodes $x$ and $y$ have an intersection of at most two nodes. 
The ball centered at node $x$ has a radius of $\tau_x-1$, and the ball centered at $y$ a radius of $\tau_y-1$. 
By construction, we know that the hop distance between $\vs$ and $x$ is $h(\vs,x)=\tau_x-1\leq \tau_x-1$ and $h(\vs,y)=\tau_y-2\leq \tau_y-1$, so $\vs$ must lie in the intersection of the two balls.
Similarly, $h(w,x)=\tau_x-2\leq \tau_x-1$ and $h(w,y)=\tau_y-1\leq \tau_y-1$, so $w$ must lie in the intersection of the two balls.
Let us assume by contradiction that there exists a third node $z$ in the intersection of these two balls. 
This implies that $h(z,x)\leq \tau_x-1$ and $h (z,y)\leq \tau_y-1$.
Either $z$ lies on the shortest path between $x$ and $y$, which we denote $P(x,y)$, or it does not.
If $z\in P(x,y)$, then it either lies to the left of $w$ or to the right of $\vs$. 
In either case, $z$ is excluded from the ball centered at $y$ or $x$, respectively. 
Thus it cannot lie on $P(x,y)$.
If $z$ does \emph{not} lie on $P(x,y)$, then there exists an alternative path $P'(x,y)\neq P(x,y)$ between $x$ and $y$ of distance at most $\tau_x + \tau_y - 2$ that contains node $z$;
this path is allowed to traverse the same node or edge multiple times.
By construction, the hop distance between $x$ and $y$ is $h(x,y) = \tau_x + \tau_y - 3$, so $P'(x,y)$ must have at least this many hops.
Moreover, since $G$ is a tree, every path between $x$ and $y$ must traverse each node in $P(x,y)$. 
Since $P(x,y)$ already contains $\tau_x + \tau_y - 3$ edges, $P'(x,y)$ should be no longer than $\tau_x+\tau_y-2$, and it should also touch an additional node $z$, 
$P'(x,y)$ must have exactly one more hop than $P(x,y)$. 
This eliminates all paths that move from node $r\in P(x,y)$ to node $z\notin P(x,y)$, then back to $r$ again.
But this implies that there exist two distinct paths between $x$ and $y$ in which no edge or vertex is traversed twice, which is a contradiction since $G$ is a tree. 
Hence there can be at most two nodes in the intersection of balls, so the probability of detection is at least 1/2 in this case. 

Combining this with the lower bound in \eqref{eq:termination}, we have
\begin{eqnarray*}
&&\prob (\mbc(\boldsymbol \tau, G) = \vs) \nonumber \\
  &\geq& \frac{\theta}{d+\theta} (1) + \frac{1}{d+\theta}\cdot \frac{1}{2} \cdot \left (1 - 2\left ( \frac{d}{d+\theta} \right )^{t-1} \right )\\
	& = & 1-\frac{d}{2(\theta +d)} - \left (\frac{d}{d+\theta}\right )^t .
\end{eqnarray*}
Since the ML estimator performs at least as well as the ball centrality estimator, the claim follows.

\subsection{Proof of Theorem \ref{thm:fs_diffusion}}
\label{sec:proof_fs_diffusion}

\begin{figure*}
\begin{eqnarray}
\prob(\tau_0 < \tmin) &=& \int_{t_0=0}^\infty \prob(\tau_0=t_0)\left [ \prob(R_1 > t_0) + \int_{r_1=0}^{t_0} \prob(R_1=r_1) \prob(\tau_1 > (t_0-r_1)) \times \right. \nonumber \\ 
								&& \left . \left [ \prob(R_2>(t_0-r_1)) + \int_{r_2=0}^{t_0-r_1}\prob(R_2=r_2)\prob(\tau_2 > (t_0-r_1-r_2)) \left [ \prob(R_3 >(t_0-r_1-r_2))+\ldots\right ]^{d-1} \right ]^{d-1} \right ]^{d_0} dt_0 \nonumber \\
& =& \int_0^\infty \lT e^{-\lT t_0} \left [ e^{-\lO t_0} + \int_0^{t_0}\lO e^{-\lO r_1}  
e^{-\lT(t_0-r_1)}\left [ e^{-\lO (t_0-r_1)} + \int_{0}^{t_0-r_1} \lO e^{-\lO r_2} e^{-\lT(t_0-r_1-r_2)} \times \right . \right . \nonumber \\
&& \left. \left. \left [  e^{-\lO (t_0-r_1-r_2)}   + \ldots \right ]^{d-1} dr_2 \right]^{d-1} dr_1 \right ]^{d_0} dt_0 \\ 
&=&  \int_0^\infty \lT e^{-\lT t_0} \left [ e^{-t_0} + \int_0^{t_0} e^{- r_1}  e^{-\lT(t_0-r_1)}\left [ e^{- (t_0-r_1)} + \int_{0}^{t_0-r_1}  e^{- r_2} e^{-\lT(t_0-r_1-r_2)} \times \right . \right . \nonumber \\
&& \left. \left. \left [  e^{- (t_0-r_1-r_2)}   + \ldots \right ]^{d-1} dr_2 \right]^{d-1} dr_1 \right ]^{d_0} dt_0 \\
&=&  \int_0^\infty \lT e^{- t_0(\lT+d_0)} \left [ 1 + e^{-t_0(\lT + d -2)}\int_0^{t_0} e^{r_1(\lT+d-2)} \left [ 1 + e^{- (t_0-r_1)(\lT + d - 2)} \int_{0}^{t_0-r_1}   e^{r_2(\lT + d-2)}  \times \right . \right . \nonumber \\ 
&& \left . \left . \left [ 1  + \ldots \right ]^{d-1} dr_2 \right]^{d-1} dr_1 \right ]^{d_0} dt_0 \label{eq:diff_fs_final}
\end{eqnarray}
\caption{Probability of detection of the first-timestamp estimator, where $d$ is the node degree. $d_0$ denotes the degree of the source, which we eventually set to $d_0=d-1$ for symmetry. We condition on the source's reporting time $\tau_0 = t_0$. $R_i$ denotes the delay of infection times between $i$'s parent and $i$.}
\label{eq:diff_fs}
\end{figure*}

To analyze the probability of detection under the first-timestamp estimator, we consider probability of the true source reporting before any other node, or $\prob(\tau_0 < \tmin)$.
We use $R_i$ to denote the random time delay between the infection times of $i$'s parent and $i$; by ``parent", we mean with respect to the infected subtree $G_t$, which is rooted at node $\vs=0$.
Moreover, we let the source have a different degree $d_0$ than the rest of tree for simplicity of calculation. 
We write the probability of detection explicitly in Figure \ref{eq:diff_fs}. 
The expression starts by conditioning on the reporting time of the true source, $\tau_0$, then conditions on the times at which other nodes receive the message. 
Recall that node $i$ receives the message at time $X_i$, and reports the message to the adversary at time $\tau_i$.

Expression \eqref{eq:diff_fs_final} can be written recursively. Define
\begin{eqnarray}
g(a) &\triangleq& e^{-a (\lT + d - 2)}\int_{0}^a e^{u(\lT+d-2)}[1+g(a-u)]^{d-1}du. \nonumber \\ 
\end{eqnarray}
Letting $s=a-u$, we get
\[
 g(a) =       \int_{0}^a e^{-s(\lT+d-2)}[1+g(s)]^{d-1}ds.
\]
Substituting $g(a)$ into Equation \eqref{eq:diff_fs_final} gives
\begin{eqnarray}
\prob(\tau_0 < \tmin) = \int_0^\infty \lT e^{-t_0(\lT + d_0)}[1+g(t_0)]^{d_0}dt_0,
\label{eq:pd_diffusion_first_spy_g}
\end{eqnarray}
where $d_0$ is the degree of the source. To make this expression symmetric with respect to the recursive function $g(\cdot)$, we assume that the source has degree $d-2$, and all other nodes have degree $d$, so $d_0=d-2$.
We subsequently compute the probability of detection by solving for $g(t_0)$, which can be written as a differential equation:
\begin{eqnarray*}
g'(a) &=& e^{-a(\lT+d-2)}[1+g(a)]^{d-1},
\end{eqnarray*}
with initial condition $g(0) = 0$.
This separable, nonlinear differential equation can be solved exactly, giving
\begin{eqnarray*}
g(a) &=& \left (\frac{(d-2)e^{-a(\lT+d-2)}+\lT}{\lT+d-2} \right )^{-\frac{1}{d-2}} - 1.
\end{eqnarray*}
Substituting into \eqref{eq:pd_diffusion_first_spy_g}, we obtain the exact probability of detection:
\begin{eqnarray*}
\prob(\tau_0 < \tmin) = \frac{\lT (\log(\lT + d - 2) - \log(\lT))}{d-2}.
\end{eqnarray*}
Letting $\lT = \theta$, we get
\begin{eqnarray}
\prob(\tau_0 < \tmin) = \frac{\theta}{d-2}\log\left ( \frac{d+\theta - 2}{\theta}\right  ),
\label{eq:pd_first_diffusion}
\end{eqnarray}
which is the final expression.

\subsection{Proof of Theorem \ref{thm:rc_diffusion}}
\label{sec:proof_rc_diffusion}

To prove this claim, we analyze the (suboptimal) reporting centrality estimator, which achieves condition \eqref{eq:const_pd}. 
Our goal is to show that under an eavesdropper adversary, reporting centrality has a strictly positive probability of detection as $d\rightarrow \infty$.

Let $R_t=\{v\in V_t | R_v(t) = 1\}$ denote the set of reporting centers at time $t$. 
We compute the probability of detection by conditioning on the event $\vs \in R_t$.
For brevity of notation, we use $\hat v$ to denote the reporting centrality source estimate in this proof.
For any fixed time $t$, we have 
\begin{align}
\prob(\mrc (\boldsymbol \tau, G) =& \vs) = 
 \underbrace{\prob(\vs\in R_t)}_{(a)} \times \nonumber \\
& \qquad  \underbrace{\prob(\mrc(\boldsymbol \tau, G) = \vs| \vs\in R_t)}_{(b)}.
\label{eq:parts}
\end{align}
Note that the probability of detection is defined for a fixed $t$.
Our goal is to lower-bound this quantity as $t\rightarrow \infty$.
The proof consists of three steps:
\begin{enumerate}
\item[$(a)$] Show that 
$
\liminf_{t\rightarrow \infty}\prob(\vs\in R_t ) \geq C_d>0.
$
\item[$(b)$] Show that 
$
\prob(\mrc(\boldsymbol \tau, G) = \vs| \vs\in R_t) = 1.
$
\item[$(c)$] Combine parts $(a)$ and $(b)$ to give the claim.
\end{enumerate}
For readability, we abbreviate our notation in this proof.  
The number of nodes in the subtree  $T^{\vs}_w$ will be denoted $N_{w}(t)$ instead of $N_{T^{\vs}_{w}}(t)$, and the number of \emph{reporting} nodes in this subtree will be denoted $Y_{w}(t)$ instead of $Y_{T^{\vs}_{w}}(t)$.
Thus $N_1(t)$ denotes the number of infected nodes in the first subtree of $\vs$, and $Y_1(t)$ denotes the number of reporting nodes in the first subtree of $\vs$.

%

\noindent \textbf{Part (a)}: Show that 
$
\liminf_{t\rightarrow \infty}\prob(\vs\in R_t ) \geq C_d.
$

\vspace{0.08in}
\noindent We demonstrate this by 
conditioning on the event that $\vs$ is the unique \emph{rumor} center in $G_t$.
This happens if and only if $\forall w\in \mathcal N(\vs)$, $N_w(t) < N(t)/2$ \cite{SZ12}, where $\mathcal N(\vs)$ denotes the neighbors of $\vs$.
We let $C_t=\{v\in V_t ~|~ v=\text{rumor center of }G_t\}$, which gives 
\begin{eqnarray*}
\prob(\vs \in R_t) \geq \underbrace{\prob(\vs \in C_t, |C_t| = 1)}_{(a_1)} \underbrace{\prob(\vs \in R_t | \vs \in C_t, |C_t| = 1)}_{(a_2)}
\end{eqnarray*}
Part $(a_1)$ is studied in Theorem 3.1 of \cite{SZ12}, which shows that
\begin{eqnarray}
\liminf_{t\rightarrow \infty} & \prob(\vs \in C_t, |C_t| = 1)  \nonumber \\
=& 1-d\left (1-I_{1/2}\left ( \frac{1}{d-2},1+\frac{1}{d-2}\right ) \right ),
\label{eq:b1}
\end{eqnarray} 
where $I_{1/2}(a,b)$ is the regularized incomplete Beta function, or the probability that a Beta random variable with parameters $a$ and $b$ takes a value in $[0,1/2)$.

For part $(a_2)$, we show that $\lim_{t\rightarrow \infty} \prob(\vs \in R_t | \vs \in C_t, |C_t| = 1)=1$.
Our approach is to first show that the fraction of reporting nodes in each tree converges almost surely to a constant as $t\rightarrow \infty$. 
We subsequently show that if $\vs$ is a unique rumor center, it is  almost surely a reporting center as $t\rightarrow \infty$.

\begin{lemma}
For all $i\in [d]$, the following condition holds as $t\rightarrow \infty$: 
\begin{equation}
\frac{Y_i(t)}{N_i(t)} \xrightarrow{a.s.} \adt = \frac{\theta }{d+\theta -2}.
\end{equation}
\label{lem:const_ratio}
\end{lemma}
(Proof in Section \ref{proof:const_ratio}) 

This lemma states that the ratio of reporting to infected nodes in each subtree converges almost surely to a constant.
The proof proceeds by describing the evolution of each subtree and the adversary's observations as a generalized P\'olya urn process with negative coefficients. 
The ratio of balls in this urn process can be shown to converge almost surely, which implies the claim.
Now we use Lemma \ref{lem:const_ratio} to show that $\vs$ is a reporting center.

Since $\vs$ is a unique rumor center, $\forall i\in [d]$ (where $[d]=\{1,2,\ldots,d\}$), $N_i(t)< N(t)/2$ \cite{SZ11a}. 
Because of this conditioning and the fact that the infected subtree sizes (normalized by the total number of infected nodes) converge almost surely \cite{SZ12}, for any outcome $\omega$ of the underlying diffusion process,
it holds that $N_i(t,\omega)/N(t,\omega)\xrightarrow{t\rightarrow \infty} (1/2 - \delta_i(\omega))$, for some $\delta_i(\omega)>0$.
Here $N_i(t,\omega)$ denotes the number of infected nodes in subtree $i$ at time $t$ under outcome $\omega$.
We wish to show that for any given, feasible set of offsets $\delta_{i}(\omega)$, $i \in [d]$, $\vs$ eventually becomes a  reporting center w.p. 1.

Since $Y_i(t)/N_i(t) \xrightarrow{a.s.} \adt $, for a given outcome $\omega$ of the underlying process, we can write $Y_i(t,\omega) =  N_i(t,\omega) (\adt + \epsilon_i(t,\omega))$, where $\epsilon_i(t,\omega) \in \reals$.
Note that $ \epsilon_i(t,\omega) \xrightarrow{t\rightarrow \infty} 0$.
We write the condition for being a reporting center as 
\begin{align}
Y_i(t,\omega) & \stackrel{?}{<}  \frac{Y(t,\omega)}{2}    \nonumber \\
 \implies &  N_i(t,\omega)  (\adt  + \epsilon_i(t,\omega))  \nonumber \\
&  \stackrel{?}{<} \frac{\adt N(t,\omega)}{2} + \frac{\sum_{j \in [d]} \epsilon_j(t,\omega)N_j(t,\omega)}{2}   \nonumber \\
\implies & N(t,\omega)  (\frac{1}{2} - \delta_i(\omega))(\adt + \epsilon_i(t,\omega) )&  \nonumber \\
		 & \stackrel{?}{<} \frac{1}{2} \left ( \adt N(t,\omega) + \sum_{j \in [d]} \epsilon_j(t,\omega) N(t,\omega) (\frac{1}{2} - \delta_j(\omega)) \right )  \nonumber \\
\implies & \epsilon_i(t,\omega) - \frac{\sum_j \epsilon_j(t,\omega)(\frac{1}{2}-\delta_j(\omega))}{1 - \frac{\delta_i(\omega)}{2}} \stackrel{?}{<} \adt \cdot \delta_{i}(\omega).\label{eq:final_condition}
\end{align}
Note that $\adt$ and $\delta_{i}(\omega)$ are both strictly positive, and 
\[
\lim_{t\rightarrow \infty} \epsilon_i(t,\omega) - \frac{\sum_j \epsilon_j(t,\omega)(\frac{1}{2}-\delta_j(\omega))}{1 - \frac{\delta_i(\omega)}{2}} = 0.
\]
Therefore, for every outcome $\omega$, there exists a time $T_\omega$ such that for all $t>T_\omega$, condition \eqref{eq:final_condition} is satisfied,
so
$
\lim_{t\rightarrow \infty} \prob(\vs \in R_t | \vs \in C_t, |C_t| = 1) = 1.
$

Putting together parts $(a_1)$ and $(a_2)$, we get
\begin{align*}
\liminf_{t\rightarrow \infty}& ~\prob(\vs\in R_t ) \\
& \geq 1-d\left (1-I_{1/2}\left ( \frac{1}{d-2},1+\frac{1}{d-2}\right ) \right ).
\end{align*}

\noindent \textbf{Part (b)}: Show that 
$
\prob(\mrc(\boldsymbol \tau, G) = \vs| \vs\in R_t) = 1.
$

\vspace{0.08in}
\noindent We show this by demonstrating that if 
$\vs$ is a reporting center, no other reporting centers exist.

%

We show that there can be at most one reporting center by contradiction. 
Suppose there are two nodes, $\vs$ and $w$, both of which are reporting centers.
Consider the $d$ subtrees adjacent to $\vs$. 
Suppose the neighbors of $\vs$ are labelled $w_1,\ldots, w_d$. 
Each of these neighbors is the root of an infected subtree $T^{\vs}_{w_i}$. 
Since $\vs$ is a reporting center,
we have two properties: 
\begin{eqnarray}
 Y_{w_i}(t) < \frac{Y(t)}{2} \qquad \forall i \in [d] \label{eq:ubound} \\
 Y_{\vs}(t) + \sum_{i \in [d]} Y_{w_i}(t) = Y(t). \label{eq:sum}
 \end{eqnarray}

Now suppose another node $u\neq \vs$ is also a reporting center. 
We label the neighbors of $u$ as $z_1,\ldots, z_d$.
We will use 
This implies that  
\begin{eqnarray}
 Y_{T^u_{z_i}}(t) < \frac{Y(t)}{2} \qquad \forall i \in [d] \label{eq:ubound2} \\
 Y_{u}(t) + \sum_{i \in [d]} Y_{T^u_{z_i}}(t) = Y(t). \label{eq:sum2}
 \label{eq:sum_node1}
 \end{eqnarray}
Suppose without loss of generality that $z_1,w_1 \in P(u,\vs)$.
In order to satisfy condition \eqref{eq:ubound2}, it must hold that 
$
Y_\vs(t) + \sum_{i =2}^d Y_{w_i}(t) < \frac{Y(t)}{2}.
$
Substituting this condition 
into \eqref{eq:sum}, and using condition \eqref{eq:ubound} implies that 
$
Y_{w_1}(t) > \frac{Y(t)}{2}. 
$
This is a contradiction because $\vs$ is assumed to be an element of $R_t$, which implies that $Y_{w_1}(t) < Y(t)/2$.
Therefore, there can be no more than one reporting center, 
so $\prob(\mrc(\boldsymbol \tau, G) = \vs| \vs\in R_t) = 1.$

\noindent \textbf{Part (c)}: Combine parts (a) and (b) to give the final claim. 

\vspace{0.08in}
\noindent  We wish to lower bound $\liminf_{t\rightarrow \infty} \prob(\mrc (\boldsymbol \tau, G) = \vs)$.
Recall from  \eqref{eq:parts} that \\
$
\prob(\mrc (\boldsymbol \tau, G) = \vs) \geq 
 \prob(\vs\in R_t ) \prob(\mrc (\boldsymbol \tau, G) = \vs| \vs\in R_t)$.
We have bounds for each term:
\begin{align*}
&\liminf_{t\rightarrow \infty}~  \prob(\vs\in R_t ) \\
&\qquad \geq 1-d\left (1-I_{1/2}\left ( \frac{1}{d-2},1+\frac{1}{d-2}\right ) \right ) \\
 &\prob(\mrc (\boldsymbol \tau, G) = \vs| \vs\in R_t) = 1,
\end{align*}
which gives
\begin{align*}
\liminf_{t\rightarrow \infty} \prob(\hv = \vs) &\geq 1-d\left (1-I_{1/2}\left ( \frac{1}{d-2},1+\frac{1}{d-2}\right ) \right ),
\end{align*}
thereby proving the claim.

\subsubsection{Proof of Lemma \ref{lem:const_ratio}}
\label{proof:const_ratio}
The evolution of the (partially unobserved) infected subgraph on regular trees can be described by a P\'olya urn process with $d$ colors of balls. 
Each ball represents one active edge between an infected and an uninfected node, 
and all active edges in the same source-adjacent subtree have the same color.
At $t=0$, we have one ball of each color, since there are $d$ active edges extending from the true source.

Because of the memorylessness of spreading, each of the active edges in $G_t$ is equally likely to spread the infection next.
The urn evolves as follows:
pick a ball uniformly at random; the subtree corresponding to the drawn color spreads the message over one of its active edges, infecting a neighbor $w$. 
Once $w$ is infected, one active edge is removed from the subtree (i.e., the edge that just spread the message), 
and $d-1$ new active edges are added (from $w$ to its uninfected neighbors).
In our urn, this corresponds to replacing the drawn ball and adding $d-2$ balls of the same color.
The replacement matrix for this urn can therefore be written as
$
A = (d-2) I_{d},
$
where $I_{d}$ denotes the $d\times d$ identity matrix.

This P\'olya urn is well-studied, and in the limit, the fraction of balls of each color is known to converge to a Dirichlet distribution \cite{mahmoud2008polya}. 
In order to model the eavesdropper's observations, we generalize the urn model by additionally giving each ball a pattern: striped or solid.
Each solid ball corresponds to an active edge in the underlying diffusion process, 
whereas each striped ball represents an active edge from an infected node to the adversary.
When such an edge fires, the adversary observes the timestamp of the sending node.

We adapt the previous urn dynamics in two key ways. 
First, when a solid ball is drawn, we still add $d-2$ solid balls of the same color, but now we also add $\theta $ striped balls of the same color.  
These represent the $\theta $ independent connections between the node and the adversary.
Second, when a striped ball is draw, we remove $\theta $ striped balls of the same color from the urn (i.e., the adversary only uses the first timestamp it receives).
Thus, the replacement matrix for a single subtree looks like
\renewcommand{\kbldelim}{(}
\renewcommand{\kbrdelim}{)}
\begin{eqnarray}
  A = \kbordermatrix{
    & \text{solid} & \text{stripe} \\
    \text{solid}  & d-2     & 0  \\
    \text{stripe} & \theta  & -\theta  
  }.
  \label{eq:rep_mtx}
\end{eqnarray}

Let $s_n$ and $r_n$ denote the number of solid and striped balls, respectively, at the $n$th draw of the urn. 
The following condition holds as $n\rightarrow \infty$: 
\begin{eqnarray}
\frac{r_n}{s_n} \xrightarrow{a.s.} \frac{\theta }{d+\theta  - 2}.
\label{eq:poly_ratio}
\end{eqnarray}
To show this, we use the following result from \cite{janson2004functional}, simplified for clarity:
\begin{customthm}{3.21 from \cite{janson2004functional}}
Consider a P\'olya urn with replacement matrix $A$. 
Assume the following conditions:
\begin{enumerate}
\item For $i,j\in [d]$, $A_{ii}\geq -1$, and $A_{ij} \geq 0$ for $i\neq j$.
\item $A_{ij} < \infty$ for all $i,j \in [d]$.
\item The largest real eigenvalue $\lambda_A$ of $A$ is positive, $\lambda_A > 0$.
\item The largest real eigenvalue $\lambda_A$ of $A$ is simple.
\item The urn starts with at least one ball of a dominating type. A dominating type is a type of ball that, when drawn, produces balls of every other type. 
\item $\lambda_A$ belongs to the dominating type.
\item The urn does not go extinct. 
\end{enumerate}
Then $n^{-1}[s_n ~~ r_n]^\intercal \xrightarrow{a.s.}\lambda_A \boldsymbol v$, where $\intercal$ denotes the transpose of a vector, $\lambda_A$ is the largest eigenvalue of replacement matrix $A$, and $\boldsymbol v$ is the corresponding right eigenvector.
\end{customthm}
Conditions 1 and 2 are satisfied by examination of $A$. 
The eigenvalues of $A$ are $(d-2)$ and $-\theta $, so conditions 3 and 4 are satisfied.
Conditions 5 and  6 are met because $\lambda_A$ belongs to the class of solid balls (i.e., a dominating type), 
and the urn starts with a solid ball by construction.
Condition 7 is met because solid balls are never removed, and we start with one solid ball.
Thus, Theorem 3.21 from \cite{janson} applies, which
 implies that $\frac{r_n}{s_n}\xrightarrow{a.s.}\frac{v_2}{v_1} = \frac{\theta }{d+\theta -2}$, since the eigenvector for $\lambda_A=(d-2)$ is $\boldsymbol v = [d+\theta -2, \theta ]$.
 
There is a one-to-one mapping between the evolution of such an urn and the spreading of the message.
Without loss of generality, we consider the evolution of the first subtree, $N_1(t)$.
Let $\beta_n$ denote the time at which the $nth$ ball is drawn. 
We define $s(t) = \max_{\{n: \beta_{n} \leq t, \beta_{n+1} > t\}} s_n $ and $r(t) = \max_{\{n: \beta_{n} \leq t, \beta_{n+1} > t\}} r_n$.
We can now map the number of reporting and infected nodes to the evolution of the P\'olya urn:
\begin{eqnarray*}
s(t) &=& 1 + (d-2)N_1(t) \\
r(t) &=& N_1(t) - Y_1(t).
\end{eqnarray*}
Solving for $Y_1(t)$ and $N_1(t)$ and taking the limit gives \newline
$
\lim_{t\rightarrow \infty} \frac{Y_1(t)}{N_1(t)} = 1 -  \frac{d-2}{\theta } \lim_{t\rightarrow \infty} \frac{r(t)}{s(t)-1} 
								= \frac{\theta }{d+\theta -2}
$
with probability 1, since both $s(t)$ and $r(t)$ tend to infinity as $t\rightarrow \infty$.

\subsection{Proof of Corollary \ref{cor:spy}}
\label{proof:spy}

The outline of this proof is similar to that of Theorem \ref{thm:rc_diffusion}.
One difference is that in the spy-based adversary, the true source $\vs$ never reports to the adversary, so $Y_\vs(t)=0$.
This does not change the proof in any significant way. 
We again condition on the event that the true source $\vs$ is a reporting center, which gives.
\begin{align}
\prob(\mrc (\boldsymbol \tau, G) =& \vs) = 
 \underbrace{\prob(\vs\in R_t)}_{(a)} \times \nonumber \\
& \qquad  \underbrace{\prob(\mrc(\boldsymbol \tau, G) = \vs| \vs\in R_t)}_{(b)}.
\end{align} 
Recall that $R_t$ is the set of reporting centers at time $t$.
For part (b), if  $\vs$ is a  reporting center, then it is a unique reporting center (from Proof \ref{sec:proof_rc_diffusion}), so the probability of detection is 1.
Thus, the key is to characterize (a), $\prob (\vs \in R_t)$, as $t\rightarrow \infty$.

As before, we lower bound this quantity by 
conditioning on the event that $\vs$ is the unique \emph{rumor} center in $G_t$.
We let $C_t=\{v\in V_t ~|~ v=\text{rumor center of }G_t\}$, which gives 
\begin{eqnarray*}
\prob(\vs \in R_t) \geq \underbrace{\prob(\vs \in C_t, |C_t| = 1)}_{(a_1)} \underbrace{\prob(\vs \in R_t | \vs \in C_t, |C_t| = 1)}_{(a_2)}
\end{eqnarray*}
We know from Proof \ref{sec:proof_rc_diffusion} that part $(a_1)$ gives
\begin{eqnarray}
\liminf_{t\rightarrow \infty} & \prob(\vs \in C_t, |C_t| = 1)  \nonumber \\
=& 1-d\left (1-I_{1/2}\left ( \frac{1}{d-2},1+\frac{1}{d-2}\right ) \right ),
\label{eq:b1}
\end{eqnarray} 
where $I_{1/2}(a,b)$ is the regularized incomplete Beta function, or the probability that a Beta random variable with parameters $a$ and $b$ takes a value in $[0,1/2)$.

For part $(a_2)$, we show that $\lim_{t\rightarrow \infty} \prob(\vs \in R_t | \vs \in C_t, |C_t| = 1)=1$.
This portion is the only real difference in proof between the present corollary and Theorem \ref{thm:rc_diffusion}.
Again, we show that the fraction of reporting nodes (or spies) in each tree converges almost surely to a constant as $t\rightarrow \infty$. 
However, unlike the eavesdropper adversary, the spy-based adversary does not require P\`olya urns to make this case. 

\begin{claim}
For all $i\in [d]$, the following condition holds as $t\rightarrow \infty$: 
\begin{equation}
\frac{Y_i(t)}{N_i(t)} \xrightarrow{a.s.} p.
\end{equation}
\label{lem:const_ratio2}
\end{claim}

This claim follows easily from the central limit theorem, since the number of spy nodes (or reporting nodes) in the $i$th subtree is simply a Binomial$(N_i(t),p)$ random variable.
Recall that $N_i(t)$ denotes the number of infected nodes in the $i$th subtree adjacent to $\vs$.
Therefore, by the same argument as Proof \ref{sec:proof_rc_diffusion}, if $\vs$ is a unique rumor center, it is  almost surely also a reporting center as $t\rightarrow \infty$.

This, in turn, implies the overall result:
\begin{equation*}
\liminf_{t\rightarrow \infty} \prob(\mrc(\boldsymbol \tau, G) = \vs) \geq C_{d} > 0.
\end{equation*}
where 
$$
C_{d} = 1-d\left (1-I_{1/2}\left ( \frac{1}{d-2},1+\frac{1}{d-2}\right ) \right ).
$$

\end{document}